\documentclass{aa}  

\usepackage[colorlinks=true, linkcolor = blue,
            urlcolor  = blue,
            citecolor = blue,
            anchorcolor = blue, bookmarks=false]{hyperref}
\usepackage{graphicx}
\usepackage{txfonts}
\usepackage{natbib}
\usepackage{enumitem} 
\usepackage{color}
\usepackage{threeparttable}
\usepackage{tikz,amsmath}
\usetikzlibrary{shapes.geometric, arrows}
\usepackage{ulem}
\usepackage{wrapfig}
\usepackage{rotating}
\usepackage{lscape} 
\tikzstyle{process} = [rectangle, minimum width=1.5em, minimum height=3.5em, text centered, draw=blue, fill=gray!10]
\tikzstyle{process2} = [rectangle, minimum width=1.5em, minimum height=3.5em, text centered, draw=white, fill=white]
\tikzstyle{arrow} = [thick,->,>=stealth]
\usepackage{cancel}
\usepackage{ulem}
\bibpunct{(}{)}{;}{f}{}{,} 

\newcommand{\teff}{$T_{\mathrm{eff}}$}

\newcommand{\logg}{\mbox{log \textit{g}}}

\begin{document}

   \title{The oldest low-$\alpha$ thin disc stars
   \thanks{XXX}
   }


   \author{Riano E. Giribaldi\inst{1}\thanks{Corresponding author: riano.escategiribaldi@inaf.it}
           \and
           Laura Magrini\inst{1}
           \and Marco Palla\inst{1}
           \and Carlos Viscasillas V{\'a}zquez\inst{1,2}
          }
   \institute{INAF – Osservatorio Astrofisico di Arcetri, Largo E. Fermi 5, 50125
   Firenze, Italy
   \and
   Institute of Theoretical Physics and Astronomy, Faculty of Physics, Vilnius University, Sauletekio av. 3, 10257 Vilnius, Lithuania
             }

    \date{Received XXX; Accepted XXX}

 
  \abstract
{The formation pathway of the $\alpha$-poor Milky Way thin disc  remains under debate. The latest observational data indicates the presence of comparable numbers of old stars in both the low- and high-$\alpha$ disc components, suggesting either coeval or overlapping formation epochs. Using GALAH DR4, we provide evidence that the $\alpha$-poor stars coexist with $\alpha$-rich stars up  to  $\sim$12.5~Gyr.
Stellar ages of  0.5~Gyr precision were re-derived for nearly  $30\,000$ subgiant  stars using accurate effective temperatures (\teff)  from the infrared flux method (IRFM) together with distances based on Gaia parallaxes. 
About 22\% of the sample constitutes a population peaking at 9.2~Gyr, with evidence for an extended old-age tail reaching 12.5~Gyr.
Among stars older than 10 and 11~Gyr, between 20 and 30\% have relative low [$\alpha$/Fe] and solar-like metallicities ($-0.5 \leq \mathrm{[Fe/H]} \leq +0.5$~dex).
These stars exhibit kinematics consistent with those of  near solar-age $\alpha$-poor stars in the thin disc. These stars can be evolved Sun-like stars, i.e. {\it old solar analogues}, whose \teff\ and gravity modified with the time. 
Comparison with a well-tested Galactic disc chemical evolution model supports the presence of a significant fraction of old, $\alpha$-poor disc stars, confirming the plausibility of our selection criteria.
We provide a catalogue with the selected sample to facilitate spectroscopic follow-ups at high resolution.}



   \maketitle
%

\section{Introduction}

Evidence for the existence of galaxies with solar or near-solar metallicities within the first $\sim$2.5~Gyr of cosmic history has steadily accumulated over the last two decades \citep[e.g.][]{maiolino2008A&A...488..463M,savaglio2012MNRAS.420..627S,sommariva2012A&A...539A.136S,scholte2025MNRAS.540.1800S,faisst2025arXiv251016106F}. These observations imply that efficient chemical enrichment can occur on relatively short timescales, leading to chemically mature stellar populations at epochs traditionally associated with predominantly metal-poor environments. If similar enrichment processes occurred in the Milky Way, they raise fundamental questions about how rapidly the chemical conditions required to produce solar-metallicity stars can be established, and hence how early environments favourable to the formation of terrestrial planets may have emerged \citep[see, e.g.][]{Spitoni2025A&A...700A..58S}. 

\begin{figure}
    \centering
    \includegraphics[width=1\linewidth]{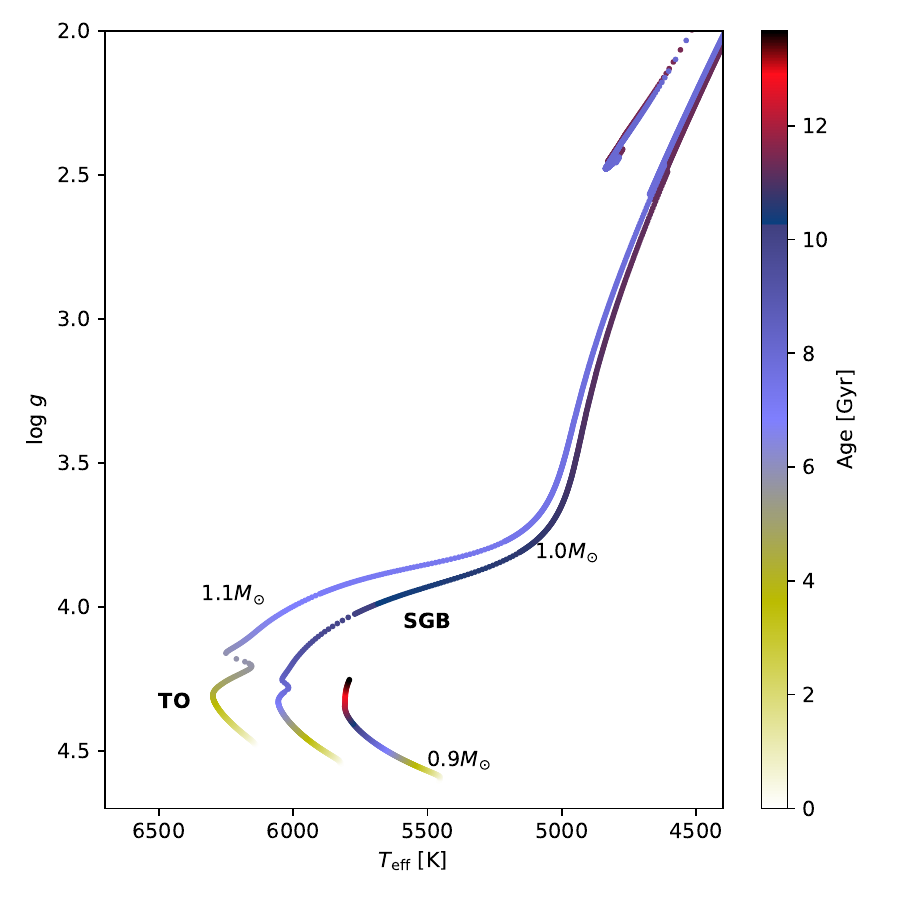}
    
    \caption{\tiny 
    Evolutionary paths of Sun like stars. BASTI evolutionary tracks \citep{hidalgo2018ApJ...856..125H} of [Fe/H] $ = -0.08$~dex and [$\alpha$/Fe] $= 0$~dex, for stars with 0.9$M_{\odot}$, 1$M_{\odot}$, and 1.1$M_{\odot}$ are shown.
    Age was truncated up to the age of the universe 13.7~Gyr \citep{plank2020A&A...641A...6P}.
     }
    \label{fig:kiel}
\end{figure}

Evidence of the existence of old metal-rich stars in the Milky Way was scarce in the past, as relatively few near-solar metallicity stars older than $\sim$10~Gyr were known prior to the \textit{Gaia} era. Before \textit{Gaia}, only a few hundred main-sequence turnoff (MSTO) and subgiant branch (SGB) stars with sufficiently precise distances and spectroscopic measurements enabling reliable age determinations were available \citep[e.g.][]{nordstrom2004A&A...418..989N,haywood2013A&A...560A.109H,Mitschang2014MNRAS.438.2753M,bensby2014A&A...562A..71B}.
These samples have been extensively used to construct and refine theoretical models of the Milky Way’s disc structure, in particular the formation of the thin and thick discs. The thin disc is characterised by relatively low [$\alpha$/Fe] abundances (hereafter $\alpha$-poor; typically defined averaging [Mg/Fe], [Si/Fe], [Ca/Fe], and [Ti/Fe]). On this basis, several studies have supported a scenario in which a thick $\alpha$-rich disc is confined to the inner Galaxy ---this is, within a Galaxy radius ($R_{\mathrm{Gal}}$) approximately lower than 9~kpc---, where the gas is progressively diluted by the accretion of more metal-poor material from the outer disc, leading to the formation of a distinct thin disc, characterised by a different chemical composition and spatial distribution \citep[e.g.][]{hayden2015ApJ...808..132H,snaith2015A&A...578A..87S,haywood2018ApJ...863..113H,haywood2019A&A...625A.105H,katz2021A&A...655A.111K}.

Nowadays, {\it Gaia} and large photometric and spectroscopic surveys are providing a whole set of information required to disentangle the origin of the Galactic stellar populations. 
Stellar ages of hundreds of thousands of MSTO and SGB stars were estimated providing a chronological timeline of the assembling history  of the Galaxy  \citep[e.g.][]{sharma2018MNRAS.473.2004S,XiangRix22,Ciuca2021, ciuca2022, giribaldi2023A&A...673A..18G, gent2024A&A...683A..74G, nepal2024A&A...688A.167N, gallart2024A&A...687A.168G}. 
MSTO and SGB stars are, indeed,  particularly valuable because their widely separated isochrones allow stellar ages to be determined with the highest precision.
In particular, \cite{beraldo2021MNRAS.502..260B,nepal2024A&A...688A.167N} and \cite{borbolato2025ApJ...994..126B} showed that the thin disc might host a significant number of stars older than $\sim$10~Gyr, including a few ones  with solar metallicity.
However, owing to their faintness, these stars have not yet been subjected to detailed spectroscopic analyses capable of probing the nucleosynthetic processes at the onset of thin-disc formation.

The uniqueness of the Sun and its planetary system relative to the broader diversity of stellar systems remains an open question, although the rapidly growing census of exoplanets is increasingly placing the Solar System in a Galactic context \citep[e.g.][]{Adibekyan2019Geosc...9..105A,Tsantaki2026arXiv260108890T}.
Establishing since when Sun-like stars started to form  is therefore crucial because it constrains the emergence of solar-metallicity conditions in the Milky Way, and therefore constrains timescales for the appearance of Earth-like planets.
In the search for stars that may resemble a newborn Sun at the earliest epochs of the Galactic thin disc, we selected a sample of very old subgiant stars, for which age determinations are among the most precise currently achievable, with kinematics consistent with thin-disc membership and metallicities close around the solar value. These objects constitute candidate \textit{old solar analogues}. Namely, whereas solar analogues simply refer to 
stars with atmospheric parameters similar to those of the Sun \citep[e.g.][]{hardorp1978A&A....63..383H,cayrel1996A&ARv...7..243C,porto2014A&A...563A..52P}, 
we call old solar analogues the stars older than $\sim$10~Gyr that  shared approximately the same parameters with the Sun at their zero age main sequence stage. 
Their existence would indicate that stars with solar-like properties were already present during the earliest phases of thin-disc formation, as suggested by recent evidence for ancient solar-metallicity thin-disc populations \citep[e.g.][]{nepal2024A&A...688A.167N,borbolato2025ApJ...994..126B}.
Such stars have evolved only slightly from their zero-age position in the Kiel diagram. For example, a solar-mass star ($1M_{\odot}$) leaves the main sequence and reaches the subgiant branch after about 10~Gyr, as Fig.~\ref{fig:kiel} shows  \citep[see also][]{bressan2012MNRAS.427..127B}, while largely preserving its initial photospheric chemical composition. Consequently, selecting  stars at the SGB region shown in Fig.~\ref{fig:kiel} provides a sample of stars with initial masses of approximately $0.9$--$1.0,M_{\odot}$, for which precise ages can be inferred from their well-separated isochrones.

In this work, we present a selection of $\alpha$-poor stars older than 10~Gyr from the GALAH~DR4 survey \citep{buder2025PASA...42...51B}, chosen as candidates for detailed chemical characterization. Our selection combines chemical and dynamical criteria. 
Surface gravities (\logg) and ages were re-derived for SGB stars using validated effective temperatures (\teff) and precise distances based on \textit{Gaia} parallaxes. Our method, which is dependent on stellar evolutionary models, was validated using the old globular cluster NGC~6352 with moderate metallicity [Fe/H] $= -0.55$ \citep{feltzing2009A&A...493..913F}, ensuring that no bias affect the inferred ages of field stars.

This paper is structured as follows. Section~\ref{sec:data} presents our data selection using GALAH catalogues. It involves our own age estimates, the method of which is described in detail. A validation of the method with the old stellar cluster NGC~6352, which stars of similar parameters to those searched in this investigation, is presented in the Appendix.
Section~\ref{sec:chars} shows the characterization of the selected old stars, where the separation of thin- and thick-disc disc stars is observed by means of chemical and orbital criteria.
Section~\ref{sec:model} shows a comparisons between the observational data with perditions of chemical evolution models.
Finally, Sect.~\ref{sec:conclusion} presents our conclusions.

\section{ Data sample}
\label{sec:data}
\subsection{Initial selection of thin disc stars}
\label{sec:1cut}

The primary goal of this work is to identify the old population of the Galactic thin disc and to derive stellar ages with the highest precision achievable through isochrone fitting, which is maximised for MSTO and SGB stars. The initial sample is based on the GALAH DR3 catalogue \citep{buder2021galah}, with subsequent improvements from GALAH DR4 \citep{buder2025PASA...42...51B}, which is built upon {\it Gaia} DR3 \citep{gaia_c} and incorporates an improved treatment of source cross-matching and binary systems. As a consequence, a number of stars classified as non-single in {\it Gaia} DR3 were removed in GALAH DR4.

GALAH DR3 includes the largest subset of stars with \teff\ derived via the infrared flux method \citep[IRFM, ][]{Blackwell1979,blackwell1980A&A....82..249B} implemented for {\it Gaia} and 2MASS photometry \cite[][]{casagrande2021}, the accuracy of which is based on the validations of  \cite{casagrande2006MNRAS.373...13C} and \cite{Casagrande2010}. 
Owing to its robust zero-point calibration and its near model-independence, IRFM \teff\ estimates together with those from interferometric angular diameters\footnote{Many of them compiled in the Gaia Benchmark Stars sample \citep{heiter2015A&A...582A..49H,soubiran2024A&A...682A.145S}.} \citep[e.g.][]{bazot2011A&A...526L...4B,karovicova2020A&A...640A..25K} and eclipsing binaries \citep[e.g.][]{miller2020MNRAS.497.2899M,maxted2022MNRAS.513.6042M} constitute primary calibration standards \citep[e.g.][]{heiter2015A&A...582A..49H,  zwintz2026arXiv260404042Z}, as \teff\ strongly influences all other stellar parameters.

Since our stellar classification relies critically on the accuracy of the age estimates, and hence on \teff, we adopt the IRFM temperatures from GALAH DR3 for the GALAH DR4 sample throughout the subsequent analysis.  
This combination of DR3 temperatures with DR4 abundances is justified by the good compatibility between the \teff\ scales of the two catalogue releases. The DR4 temperatures are calibrated against the Gaia Benchmark Stars \citep{heiter2015A&A...582A..49H}, whose temperature scale is consistent with the IRFM scale \citep{giribaldi2021A&A...650A.194G,giribaldi2023A&A...679A.110G}. Thus, combining the DR3 IRFM \teff\ scale with the DR4 chemical abundances should not introduce a significant systematic inconsistency. At the same time, DR4 provides more precise elemental abundances than DR3, owing to the larger number of spectral lines used in their determination. We therefore adopt the DR3 IRFM temperatures together with the DR4 abundances throughout our analysis, thereby retaining the well-established IRFM \teff\ scale while taking advantage of the improved precision of the DR4 chemical abundances.
Appendix~\ref{sec:compat_params} shows the compatibility of the \teff\ and \logg\ scales related to the ages in this work with those of GALAH~DR4.

We further restrict the sample to disc kinematics by selecting stars located in the prograde region of the Lindblad diagram (orbital energy versus angular momentum $L_Z$), following the global separation illustrated in Fig.~4 of \citet{giribaldi2023A&A...673A..18G}.
We then impose quality cuts on the atmospheric parameters and parallaxes, requiring \texttt{flag\_fe\_h = 0}, \texttt{flag\_alpha\_fe = 0}, and \texttt{flag\_sp = 0}, 
and selecting stars with $\varpi \geq 0.7$ mas and $\varpi / \sigma(\varpi) > 5$ to ensure the determination of reliable distances. To focus on MSTO and SBG stars, we further apply $T_{\mathrm{eff}} < 6750$~K to exclude hot young stars, $\log g < 4.4$ dex to maintain the regime where our isochrone-based gravity estimates remain reliable, and $-1.5 < \mathrm{[Fe/H]} < +0.45$ dex. These criteria result in a sample of $166\,441$ stars, spanning a wide range of ages and covering the full metallicity distribution of the Galactic thin disc.

\subsection{Stellar age determination}
\label{sec:ages}

We derived stellar ages using the Python code $q^2$ \citep{Ramirez2014}, which performs a frequentist inference framework \citep[as implemented in][]{ramirez2013ApJ...764...78R} based on the input quantities \teff, [Fe/H], extinction-corrected $V$ band magnitude ($V_0$), and parallax $\varpi$,  along with their respective uncertainties. The code inverts the parallax to compute distances and derive absolute magnitudes ($M_V$), which provide a good approximation for nearby stars. For $\varpi \lesssim 1.3$~mas, this approximation may introduce small biases; therefore, we instead adopt the geometric distances from \citet{bailer-jones2021AJ....161..147B}, which include corrections for the Gaia parallax zero-point offset \citep{2021A&A...649A...4L}.

The code employs Yale--Yonsei isochrones \citep{kim2002,yi2003} of solar-scaled composition, sampled in metallicity with 0.02~dex spacing. To account for $\alpha$-enhancement effects, we follow the approach of \cite{spina2018MNRAS.474.2580S}, which maps [$\alpha$/Fe] into an effective metallicity using the prescription of \cite{salaris1993ApJ...414..580S}:
\begin{equation}
\label{eq:alpha_en}
[\mathrm{M/H}] = [\mathrm{Fe/H}] + \log(0.638 \times 10^{[\alpha/\mathrm{Fe}]} + 0.362)
\end{equation}

\noindent  where the average of [Mg/Fe], [Si/Fe], [Ca/Fe], and [Ti/Fe] is adopted as [$\alpha$/Fe]. As shown by \cite{spina2018MNRAS.474.2580S,XiangRix22} and \cite{vazquez2026}, this correction avoids systematic age biases of 1–2~Gyr that mainly affect stars older than $\sim$8~Gyr. We also apply a zero-point correction of +0.04~dex (i.e. an offset of $-0.04$~dex is assumed) in [M/H] to account for the effects of atomic diffusion at solar parameters \citep{melendez2012A&A...543A..29M,dotter2017ApJ...840...99D}, which allows the recovery of the solar age of 4.6~Gyr \citep{connelly2008ApJ...675L.121C,amelin2010E&PSL.300..343A} inferred from fundamental parameter constraints.

The method is validated in Appendix~\ref{sec:validation} with the old globular cluster NGC~6352, while its sources of uncertainty and potential biases are examined in Appendix~\ref{sec:deviation}. 
The method returns posterior probability distributions for \logg, age, mass, luminosity, and radius.  Our analysis indicates that the age uncertainties are dominated by the combined effects of \teff\ and $M_V$ uncertainties, with the latter incorporating contributions from the photometric magnitude, reddening, and parallax uncertainties.  The age precision improves significantly for SGB stars compared with TO stars, as the sensitivity of the derived age to deviations in \teff\ from its true value becomes negligible at this evolutionary stage. A detailed analysis of these effects is presented in Appendix~\ref{sec:deviation}.

Uncertainties in IRFM \teff\ are assumed to be 70~K, obtained by combining in quadrature the intrinsic scatter and zero-point uncertainty of the colour--\teff\ calibrations (60~K and 20~K, respectively; \citealt{casagrande2021}). 
Parallax uncertainty is adopted from the Gaia catalogue, which for our stars is statistically equivalent to to that obtained by propagating the uncertainty in the geometric distance.
The uncertainty of $V_0$  is dominated by two sources that were added in quadrature.  These are, the uncertainty of the Gaia-Johnson transformation given in Eq.~\ref{eq:gaia_johnson} ($\sigma(V_{calib}) = 0.03$~mag) and the uncertainty of extinction correction. For the latter, the uncertainty given by Eq.~\ref{eq:reddening_error} is scaled to the $V$ band using the standard coefficient $R_V = 3.1$.
A robust assessment of the reddening uncertainties is essential for evaluating their contribution to the age error budget and identifying potential systematic biases. This analysis is presented in Appendix~\ref{sec:redd}, where we find that any systematic effects associated with the adopted reddening, if present, are expected to become significant primarily for stars in regions with $E(B-V) \gtrsim 0.2$.
Such effects would concern only a marginal fraction (5-10\%) of the stars older than 10, 11, and 12~Gyr, and would bias the inferred ages towards younger or older values if $E(B-V)$ were underestimated or overestimated, respectively.

\subsection{Selection of the oldest disc populations}
\label{sec:sel_old}

\begin{figure*}
    \centering
    \includegraphics[width=0.9\linewidth]{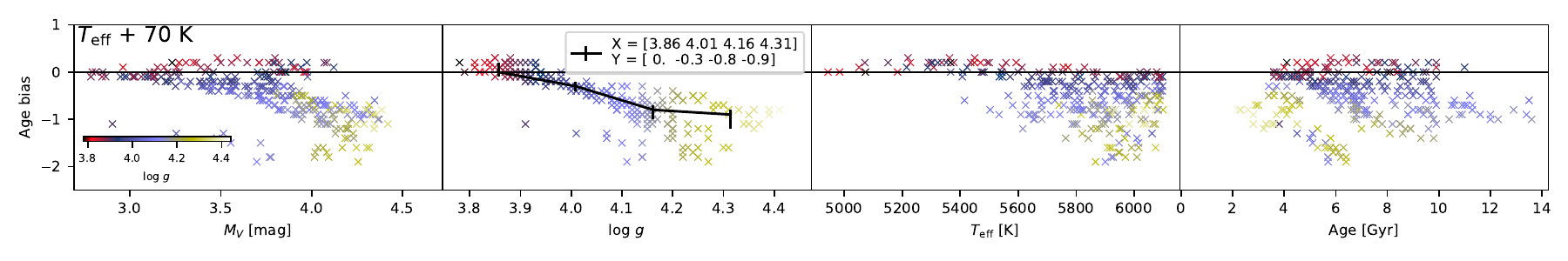}
    \includegraphics[width=0.9\linewidth]{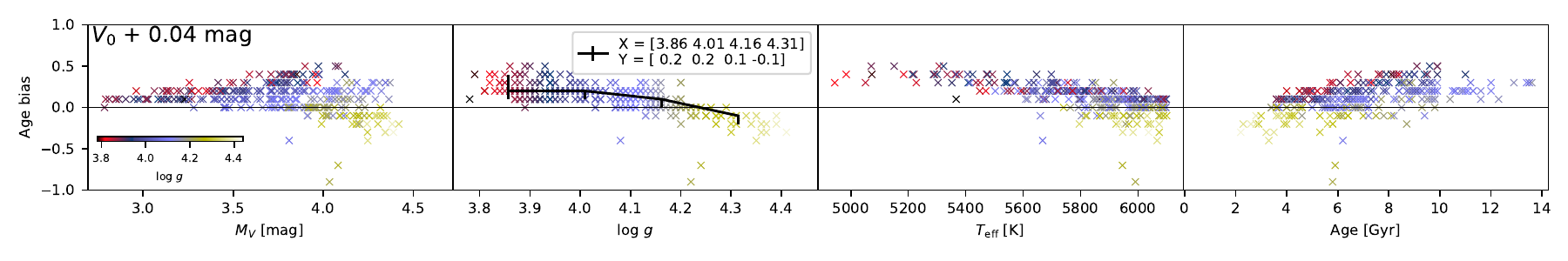}
    \includegraphics[width=0.9\linewidth]{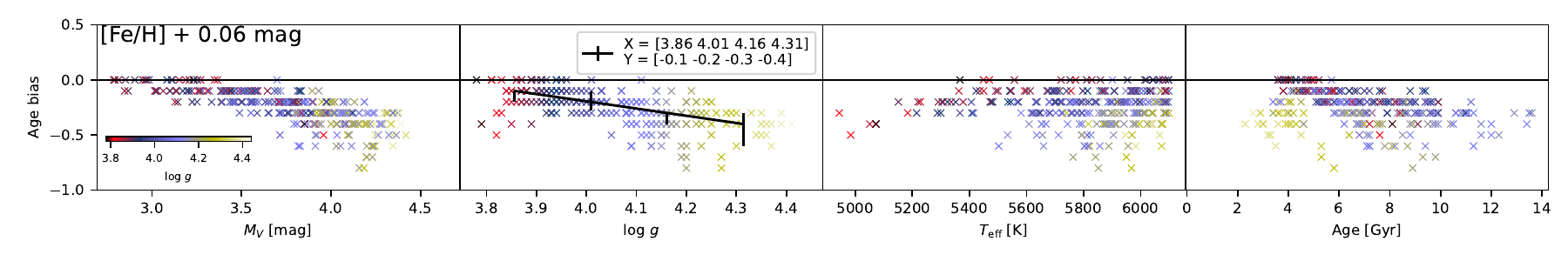}
    \includegraphics[width=0.9\linewidth]{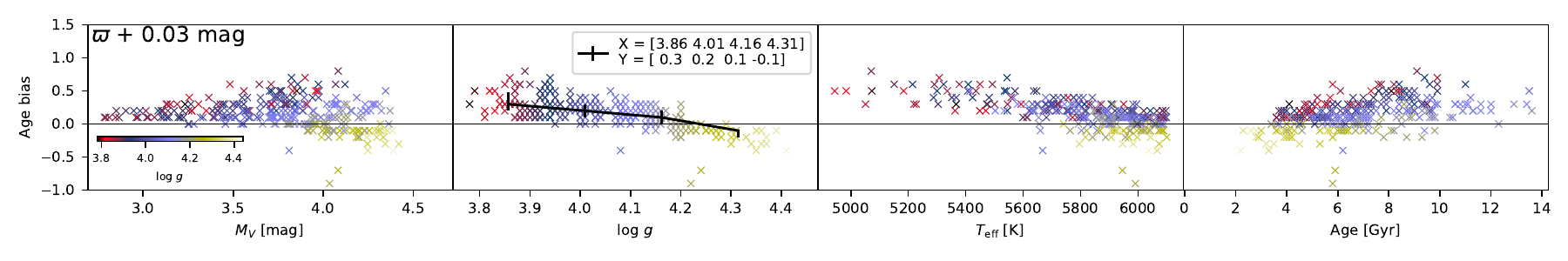}
    \caption{\tiny  
    Correlation between age biases and relevant stellar parameters. From top to bottom, the panels show the response of the age biases to typical variations in \teff, $V_0$, [Fe/H], and parallax, as indicated in the left column. In the panels showing the age bias as a function of \logg, the trends are defined by median values computed in equally spaced bins and connected by lines; corresponding quantities are noted in the legends. 
    The sample used correspond to 300 stars randomly selected from the sample of $52\,815$ stars, see main text.
    }
    \label{fig:map_error}
\end{figure*}

We restricted the sample of $166\,441$ stars to those with \teff\ $<6100$~K. This limit corresponds approximately to the main-sequence turn-off temperature of a 5-Gyr population at solar  metallicity and therefore excludes younger, more massive stars. 
The algorithm  for age determination  processed data of 91425 stars and converged for $90\,099$. Of these, $52\,815$ have  internal\footnote{Uncertainty outcome of the statistical method.} age uncertainties smaller than 1.5~Gyr; this corresponds to both the 16th and 84th percentiles of their age posterior distributions ($\sigma$-like errors henceforth).
Figure~\ref{fig:map_error} illustrates the changes in the inferred ages induced by typical variations in \teff, $V_0$, [Fe/H], and parallax.

\begin{figure}
    \centering
    \includegraphics[width=1\linewidth]{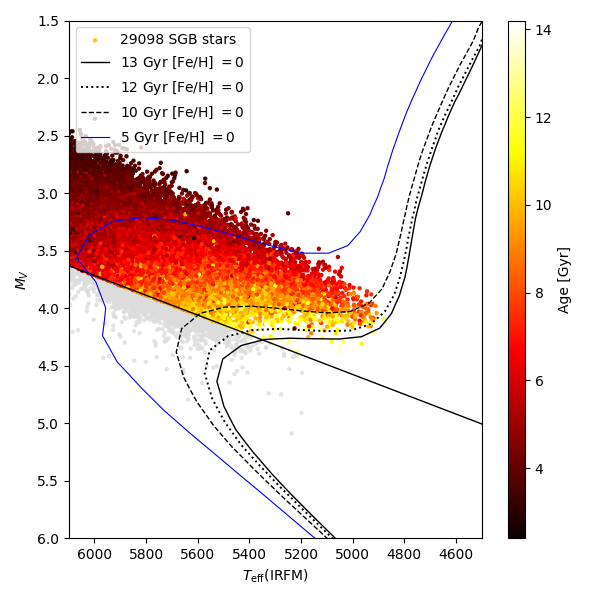}
    
    \caption{\tiny 
    $M_V$--\teff\ diagram of the $35\,879$ field stars with \logg\ $<4.1$~dex. The $32\,632$ stars retained for the age analysis, corresponding to the SGB selection, are colour-coded by age. Stars excluded by the relation $M_V \geq -0.00086$\teff\ $+ 8.876$ (black line) are shown in grey.
    Yale-Yonsei isochrones with ages and metallicities according to the labels are shown as reference.
     }
    \label{fig:Mv_teff}
\end{figure}

To minimise spurious age overestimates, we applied the prescription established in Appendix~\ref{sec:deviation}. 
This is, we first retained stars with \logg\ $\leq 4.1$~dex ($30\,972$ stars), and subsequently selected only those located within the SGB region of the $M_V$--\teff\ diagram.
Figure~\ref{fig:Mv_teff} illustrates the latter restriction, where an empirical cut is defined.
The colour-code clearly exhibits  stars older than 10~Gyr on the SGB traced by the isochrones.  This cut retained $29\,098$ stars; we refer to this subset as the {\it characterised sample}, which forms the basis of the analysis below.
Figure~\ref{fig:map_error} shows that, for these stars, the combined effect of typical uncertainties in the input parameters leads to age uncertainties of $\sim$0.5~Gyr.
Appendix~\ref{sec:ages_C} compares our age estimates with the catalogued ones.

\begin{figure}
    \centering
    \includegraphics[width=1\linewidth]{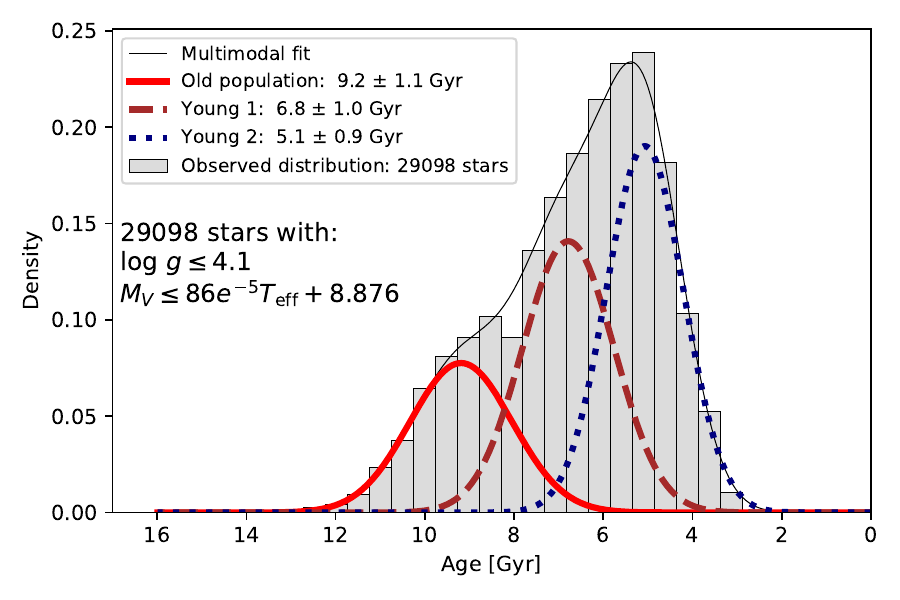}\\
    \includegraphics[width=1\linewidth]{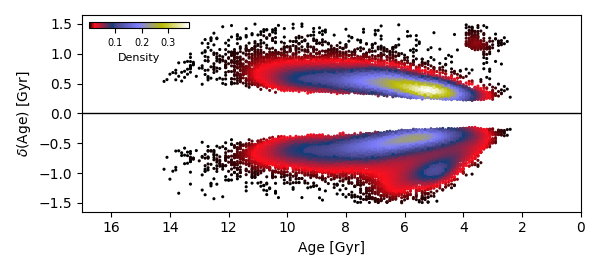}
    
    \caption{\tiny 
     {\it Top panel:} Age normalised histogram of the characterised sample. The age axis is segmented into bins of 0.48~Gyr. 
     A multimodal fit is displayed by the black thin line, the individual distributions of which are represented according to the legends.
     Each population is labelled by its nickname. Their medians and corresponding $\sigma$ dispersions are also indicated.
     {\it Bottom panel: } $\sigma$-like errors of the ages of the stars that compose the histogram on the top panel.
     }
    \label{fig:histo_eye}
\end{figure}

Top panel in Fig.~\ref{fig:histo_eye} shows the normalised age distribution of the characterised sample, while the bottom panel shows the distribution of the star-by-star $\sigma$-like errors. 
A multimodal fit reveals a prominent population centred at 9.2~Gyr, whose Gaussian component comprises 22.31\% of the sample ($6492$ stars) and extends to $\approx$12.5~Gyr at 3$\sigma$. 
Two additional components are identified at younger ages, centred at 6.8~Gyr ({\it Young~1}) and 5.1~Gyr ({\it Young~2}), typically associated to the thin disc in the literature. 

To evaluate the robustness of the oldest tail of the age distribution given the individual age uncertainties, we estimated the expected contribution from stars scattered to older ages.
For each star with a measured age below a given threshold ($\tau_{up}$), we used Monte Carlo simulations to sample its age posterior, adopting a Gaussian distribution with an upper uncertainty corresponding to the 84th percentile, and calculated the probability of its age exceeding $\tau_{up}$. Summing these probabilities provides an estimate of the expected number of stars scattered above
each threshold.
Table~\ref{tab:contamination} shows that this contribution becomes increasingly important toward older ages, closely approaching $\sim$100\% for beyond $ 13$~Gyr. 
The closer the values are to 100\%, the better the number of stars above a given $\tau_{up}$ is explained by the dispersion due to the age precision.
For stars older than 11.5, 12, and 12.5~Gyr, the corresponding ratios are  $\sim$75\%.
Thus, the expected upward scattering accounts for most, but not all, of the stars above these age thresholds, with the remaining $\sim$25\% suggesting the presence of a genuine old population.
Above 13~Gyr, the expected contribution of 32 stars is consistent with the entire observed sample of 31 stars. 

\begin{table}
\caption{Expected contribution of age uncertainties to the old-age tail}
\label{tab:contamination}
\centering
\tiny 
\begin{threeparttable}
\begin{tabular}{l|cccc}
\hline\hline
$\tau_{up}$  & Observed  & Expected upward-scattered & Expected/observed \\
(Gyr) & stars & stars & (\%) \\
\hline
10.0 & 1690 & 496 & 29\\
10.5 & 879 & 398 & 45\\
11.0 & 452 & 262 & 58\\
11.5 & 212 & 163 & 77\\
12.0 & 112 & 92 & 82\\
12.5 & 68 & 50 & 73\\
13.0 & 31 & 32 & 105\\
\hline
\end{tabular}
\begin{tablenotes}
\item{} Notes. First column lists the age threshold. Second column displays the number of stars in the characterised sample with ages over the threshold. Third column lists the number of stars scattered over the threshold computed by Monte Carlo simulations. Last column lists percentage of potential age contaminants over the threshold.
\end{tablenotes}
\end{threeparttable}
\end{table}


\section{Characterization of the data sample}
\label{sec:chars}

Using the chemical abundances and orbital properties from GALAH DR4, we characterised our selected sample to separate thin- and thick-disc stellar populations based on their combined kinematic and chemical properties.

\subsection{Chemical properties}
\label{sec:chem}

\begin{figure}
    \centering
    \includegraphics[width=1\linewidth]{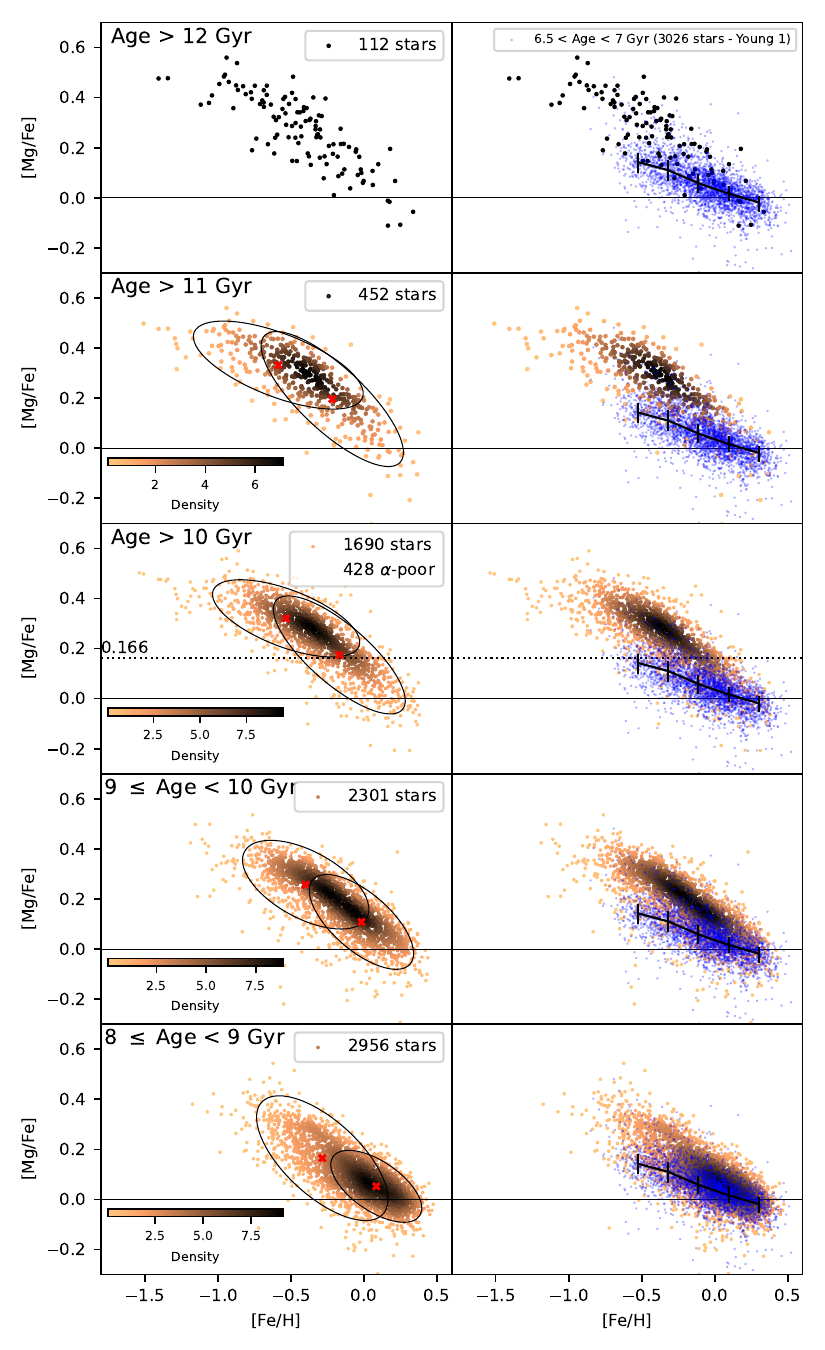}
    \caption{\tiny 
     [Mg/Fe] versus [Fe/H] diagrams. 
     Each row exhibits old stars of three different age ranges, colour-coded according to their probability density scaled in the bars.
     \textit{Left panels:} Ellipses denote the 2$\sigma$ ($\sim$95\% probability) contours of the 2D Gaussian-mixture models fitted to the data distributions; red crosses mark their centres. The dotted lines indicate the lower boundaries of the high-$\alpha$ population ellipses.
     \textit{Right panels:} Same distributions as in left panels are shown. Blue dots represent stars of the \textit{Young~1} population (cut of $6.5 \leq$ age $\leq 7$~Gyr) in Fig.~\ref{fig:histo_eye}.
     Black lines connect medians of [Mg/Fe] of the blue dots computed in equally separated bins. Vertical bars represent 25 and 75\% quartiles around the medians.    
     }
\label{fig:MgFe}
\end{figure}

\begin{figure}
    \centering
    \includegraphics[width=1\linewidth]{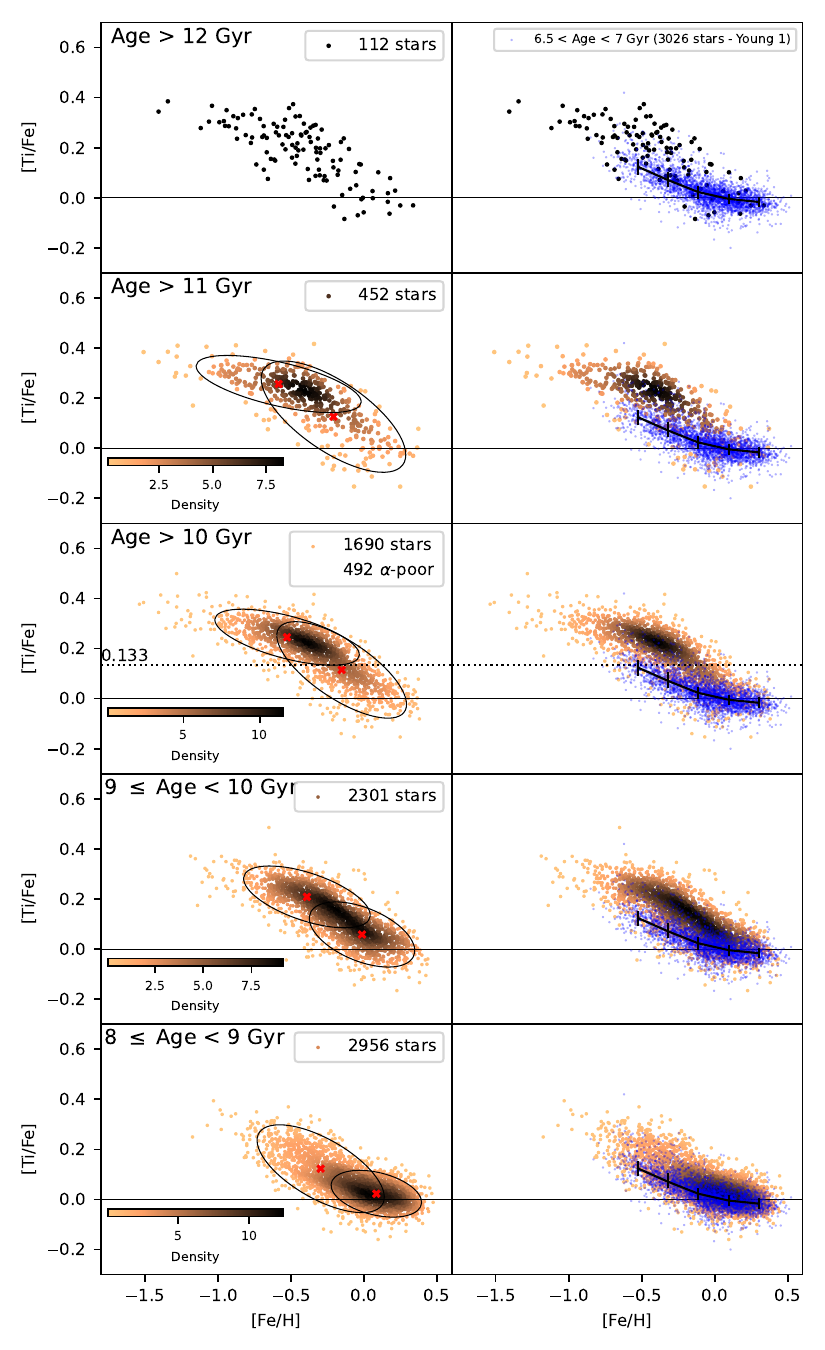}
    \caption{\tiny 
    [Ti/Fe] versus [Fe/H] diagrams. The elements of the plots are the same as in Fig.~\ref{fig:MgFe}.     
     }
\label{fig:TiFe}
\end{figure}

We identified stellar populations belonging to the $\alpha$-poor and $\alpha$-rich discs using the most sensitive chemical discriminants, [Mg/Fe] and [Ti/Fe], while we find that other $\alpha$-elements such as Si and Ca are less effective, as commonly reported in the literature \citep[e.g.][]{nissen2010A&A...511L..10N,vincenzo2021MNRAS.508.5903V,buder2025PASA...42...51B}.
Figures~\ref{fig:MgFe} and \ref{fig:TiFe} show the oldest stars in the sample in the [Mg/Fe]--[Fe/H] and [Ti/Fe]--[Fe/H] planes, where we grouped them into age intervals. The right-hand panels include thin-disc stars with solar ages (Young~1 population; blue symbols) to highlight their overlap with the old low-$\alpha$ population. The corresponding abundance trends are represented using the 16th, 50th, and 84th percentiles.

The three oldest age intervals show the coexistence of both 
$\alpha$-rich ([Mg/Fe] $\gtrsim 0.2$~dex) and $\alpha$-poor ([Mg/Fe] $\lesssim 0.2$~dex) 
populations with comparable numbers of stars, even at ages exceeding  11~Gyr. The $\alpha$-poor stars overlap with the sequence of solar-age thin disc stars at solar metallicity regardless their ages.
The presence of such old, chemically thin-disc-like stars is consistent with scenarios in which the thin and thick discs formed contemporaneously during the early evolution of the Milky Way \citep[e.g.][]{grisoni2026arXiv260510596G}. Furthermore, the existence of $\alpha$-poor stars older  than  11~Gyr with near-solar metallicities is in line with observations of massive galaxies already reaching near-solar metallicities at redshifts $z \gtrsim 3$ \citep[e.g.][]{maiolino2008A&A...488..463M, savaglio2012MNRAS.420..627S, sommariva2012A&A...539A.136S, scholte2025MNRAS.540.1800S, faisst2025arXiv251016106F}.

To select the stars with the highest probability of being $\alpha$-poor, we applied Gaussian mixture models \citep{pedregosa2011JMLR...12.2825P}.  The abundance distributions of stars older than 10 and 11~Gyr closely resemble. In both cases, the region most likely associated with $\alpha$-poor stars lies below the lower boundary of the $\alpha$-rich component, as defined by the Gaussian ellipses (horizontal dotted lines). These boundaries  in both the [Mg/Fe] and [Ti/Fe] planes are nearly consistent and align with the upper envelope of the solar-age thin-disc sequence in blue. 
The total number of stars and the $\alpha$-poor counts  for age~$>10$~Gyr are reported in the figures. 
We classify as old $\alpha$-poor (old $\alpha$-rich) the stars that consistently lie below (above) the dotted boundaries in the third-row panels of Figs.~\ref{fig:MgFe} and \ref{fig:TiFe}.
Out of  1690 stars older than 10~Gyr,  364  are classified as $\alpha$-poor and  1134 as $\alpha$-rich, while  192 stars show ambiguous classifications and are therefore labelled as transition stars. These numbers imply an overall $\alpha$-poor fraction of approximately  20-30\%. The bottom panels of both figures further show that the $\alpha$-poor population begins to dominate between 8 and 9~Gyr, suggesting a transition in the dominant chemical regime of the thin disc at $\sim$9.5~Gyr.

\subsection{Orbital properties}

\begin{figure*}[!htbp]
    \centering
    \includegraphics[width=0.32\linewidth]{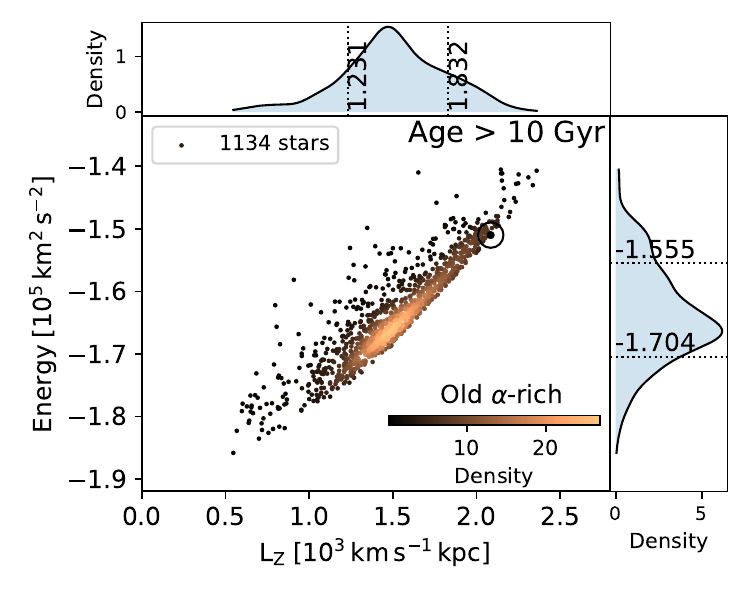}
    \includegraphics[width=0.32\linewidth]{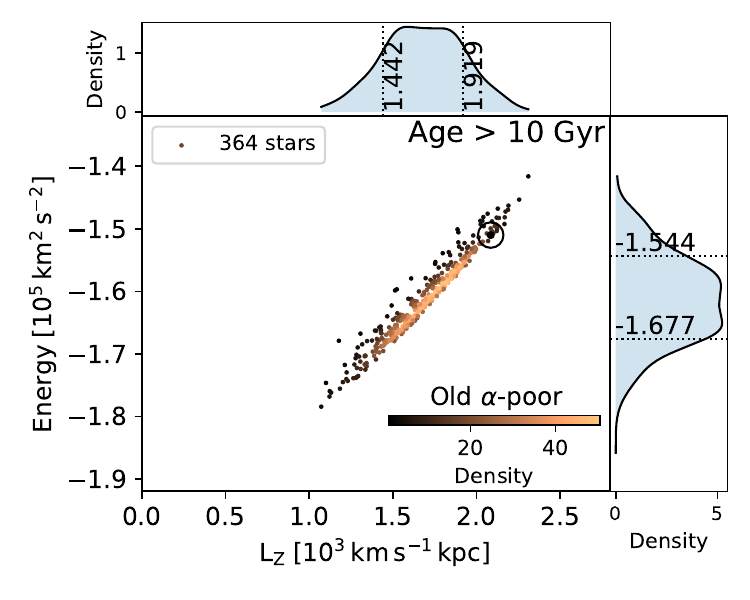 }
    \includegraphics[width=0.32\linewidth]{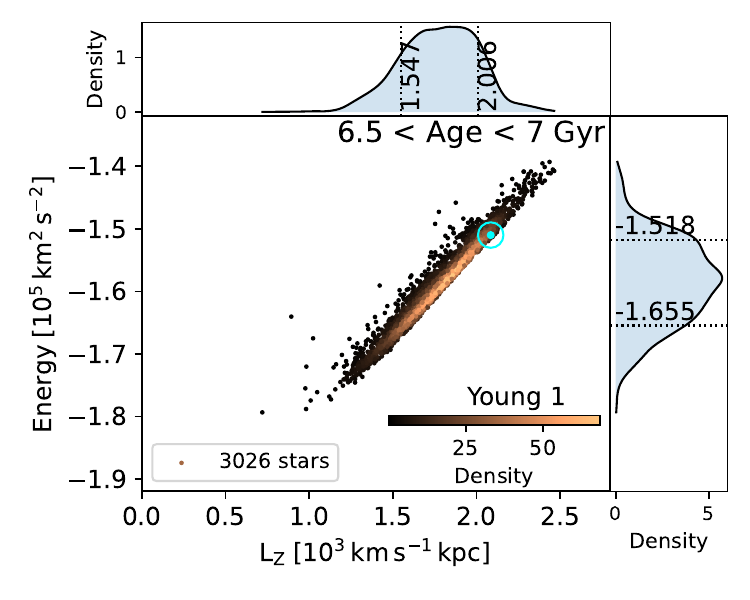}\\
    \includegraphics[width=0.32\linewidth]{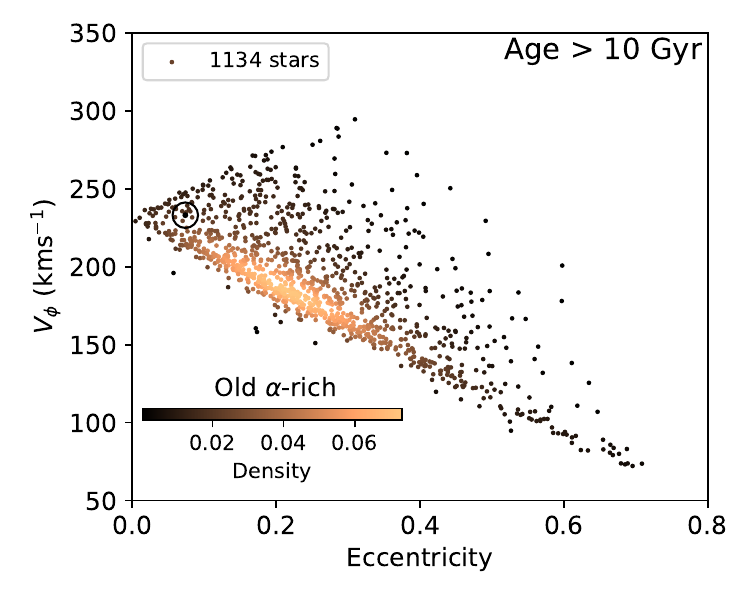}
    \includegraphics[width=0.32\linewidth]{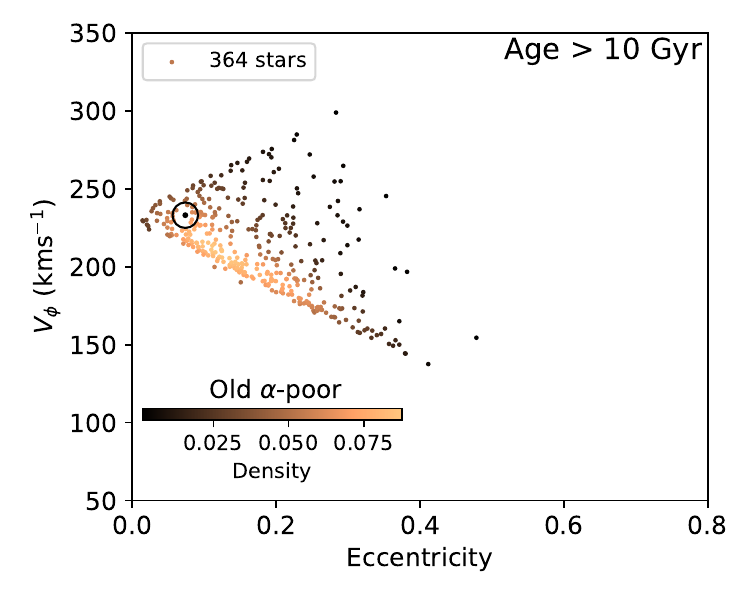}
    \includegraphics[width=0.32\linewidth]{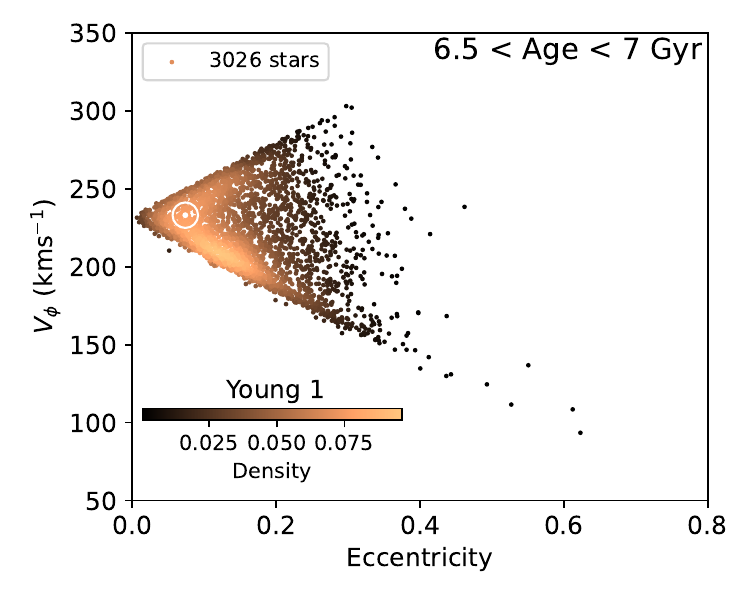}
    \caption{\tiny Dynamical properties of the old and young populations.
     {\it Top panels: } Lindblad diagrams for $\alpha$-rich stars older than 10~Gyr (left panel), $\alpha$-poor stars older than 10~Gyr (centre panel), and the solar-age stars ($5.1 \leq \mathrm{age} \leq 5.5$~Gyr) of the \textit{Young~1} population in Fig.~\ref{fig:histo_eye} (right panel); i.e. blue dots in Figs.~\ref{fig:MgFe} and \ref{fig:TiFe}. Point colours follow the corresponding colour bars. Marginal density distributions are shown above and to the right of each diagram; dotted lines indicate the 16th and 84th percentiles. The solar position is marked by $\odot$.
     {\it Bottom panels: } Rotational velocity ($V_\phi$) versus eccentricity. Symbols are the same as in the top panels.}
    \label{fig:lind}
\end{figure*}

We examined the orbital properties of the old $\alpha$-rich and $\alpha$-poor populations identified in the previous section using the kinematic and dynamical information provided by the GALAH DR4 catalogue.
The top-left and top-centre panels of Fig.~\ref{fig:lind} show the distributions of the $\alpha$-rich and $\alpha$-poor populations in the Lindblad diagram for stars older than 10~Gyr, while  6.5-7~Gyr stars from the Young~1 population are shown in the top-right panel as a reference for the typical thin-disc kinematics.
The old $\alpha$-rich population exhibits a widely dispersed distribution, consistent with thick-disc kinematics \citep[e.g.][]{li2018ApJ...860...53L}. In contrast, the old $\alpha$-poor population shows a narrow distribution concentrated in the same energy--$L_Z$ region as the thin disc, as indicated by the 16th and 84th percentiles of the density distributions.
These contrasting kinematic properties demonstrate that old $\alpha$-poor stars are  virtually dynamically indistinguishable from the thin disc, while the old $\alpha$-rich population is consistent with a thick-disc origin, highlighting that chemical separation remains reflected in orbital structure even at the oldest ages probed here.

\section{Comparison with Galactic Chemical Evolution Models}
\label{sec:model}

The presence of a significant fraction ($\sim$25\%) of stars older than 10 Gyr with low-$\alpha$ element abundance opens the question whether current chemical evolution scenarios can predict such stars. For this reason, we consider the two-infall model proposed by \citet{Palla22}. 


In brief, the model assumes that the Milky Way (MW) disc forms by means of two sequential gas accretion episodes, separated by a 3.25 Gyr delay, normally associated to the the chemical thick and thin disk, respectively, relative to the first event.
Within the adopted framework, stellar radial migration is also implemented to take into account the observed effect of radial mixing of stellar population born at different radii (see, e.g. \citealt{Kordopatis15,Feltzing20}). In particular, the authors followed the parametric implementation described by \citet[][see Eq. (8) in \citealt{Palla22}]{Frankel18}. For all the information and details about the adopted model prescriptions and assumptions, we address the reader to \citet{Palla22}. To perform a fair comparison between the model predictions and the selected old population within GALAH, we take into account the cuts in \teff\, and \logg\, by applying PARSEC (v1.2s) + COLIBRI (s37) tracks
.
Moreover, we took into account uncertainties in stellar age and abundance determinations by building a "synthetic" chemical evolution model, namely a mock stellar catalogue of surviving stars. In particular, we add at each model timestep a random error to the ages and the [Fe/H] and [Mg/Fe] abundances of the stars, having a Gaussian distribution with standard deviation as typical uncertainties of stars.

\begin{figure}
    \centering
    \includegraphics[width=1.\linewidth]{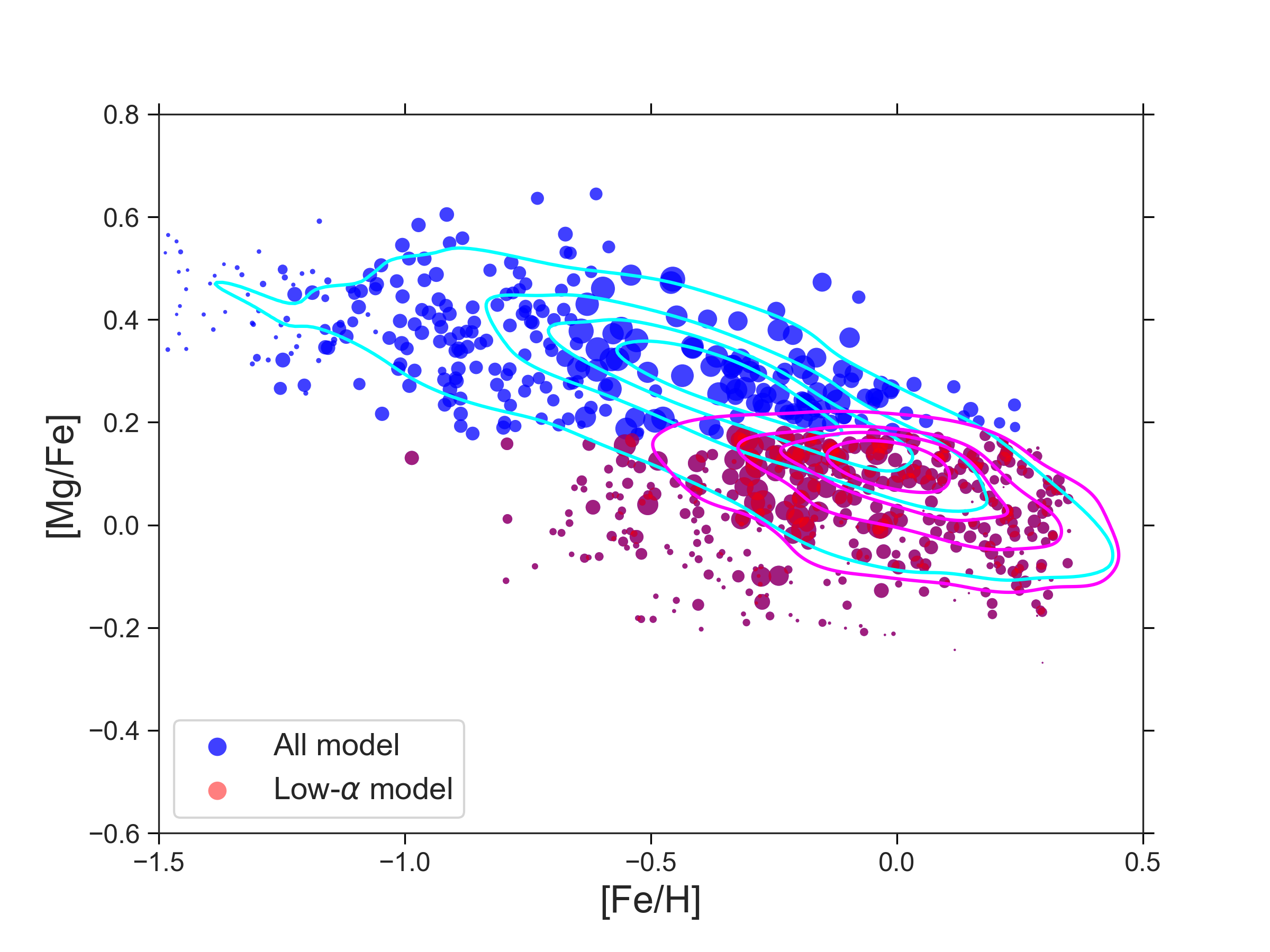}
    \caption{\tiny 
      Two-infall model predictions for [Mg/Fe] vs. [Fe/H]. Coloured points are the synthetic model predictions for stars older than 10 Gyr, with red points being compatible with the old-$\alpha$ poor population defined in Section \ref{sec:chem}. Point size is proportional to the number of stars formed in the model timestep. Cyan and magenta contours represent density lines of the full sample of stars older than 10 Gyr and the $\alpha$-poor population older than 10 Gyr.}
    \label{fig:model}
\end{figure}

The result of the comparison between the two-infall framework and the data presented in this work is displayed in Fig. \ref{fig:model}. We show the [Mg/Fe] vs. [Fe/H] "synthetic" model predictions and data for observed stars older than 10 Gyr\footnote{We do not perform as data-model comparison for [Ti/Fe] vs. [Fe/H], as due to the long-standing problem in stellar nucleosynthesis to obtain reliable yields for Ti (see \citealt{Romano10,Prantzos18,Pepe25}).}.  The figure shows that  in this scenario, it is indeed possible to have $\alpha$-poor stars with disc-like properties that are very old, making our selection of old $\alpha$-poor candidates fully plausible.  In addition, predicted number counts for  the low-$\alpha$ population (red points in Fig. \ref{fig:model}) represent around 40\% of the entire model population older than 10 Gyr, a value larger but still in line than the $\sim$20-30\% shown by observations. 

\begin{figure*}
    \centering
    \includegraphics[width=0.495\textwidth]{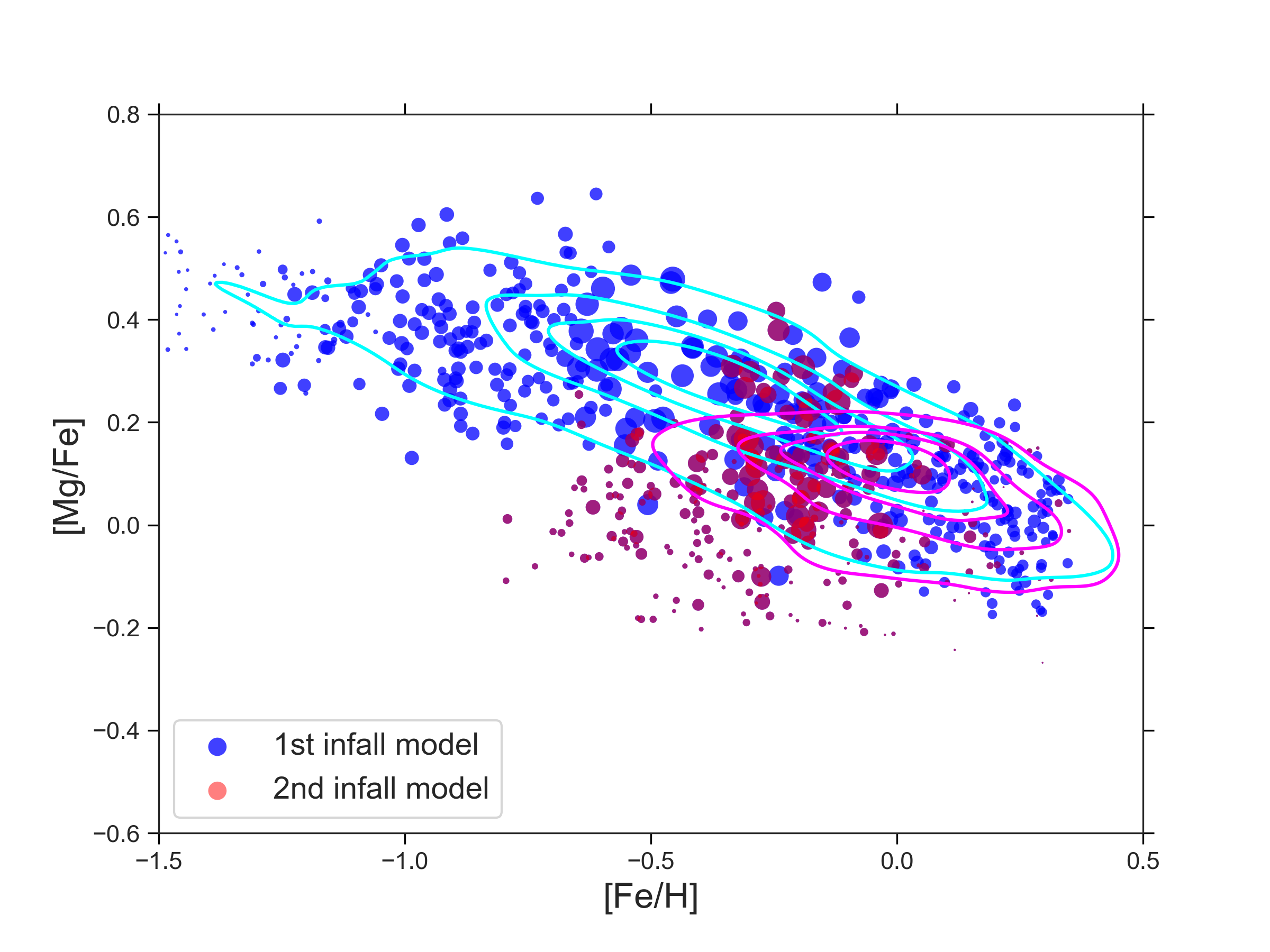}
    \includegraphics[width=0.495\textwidth]{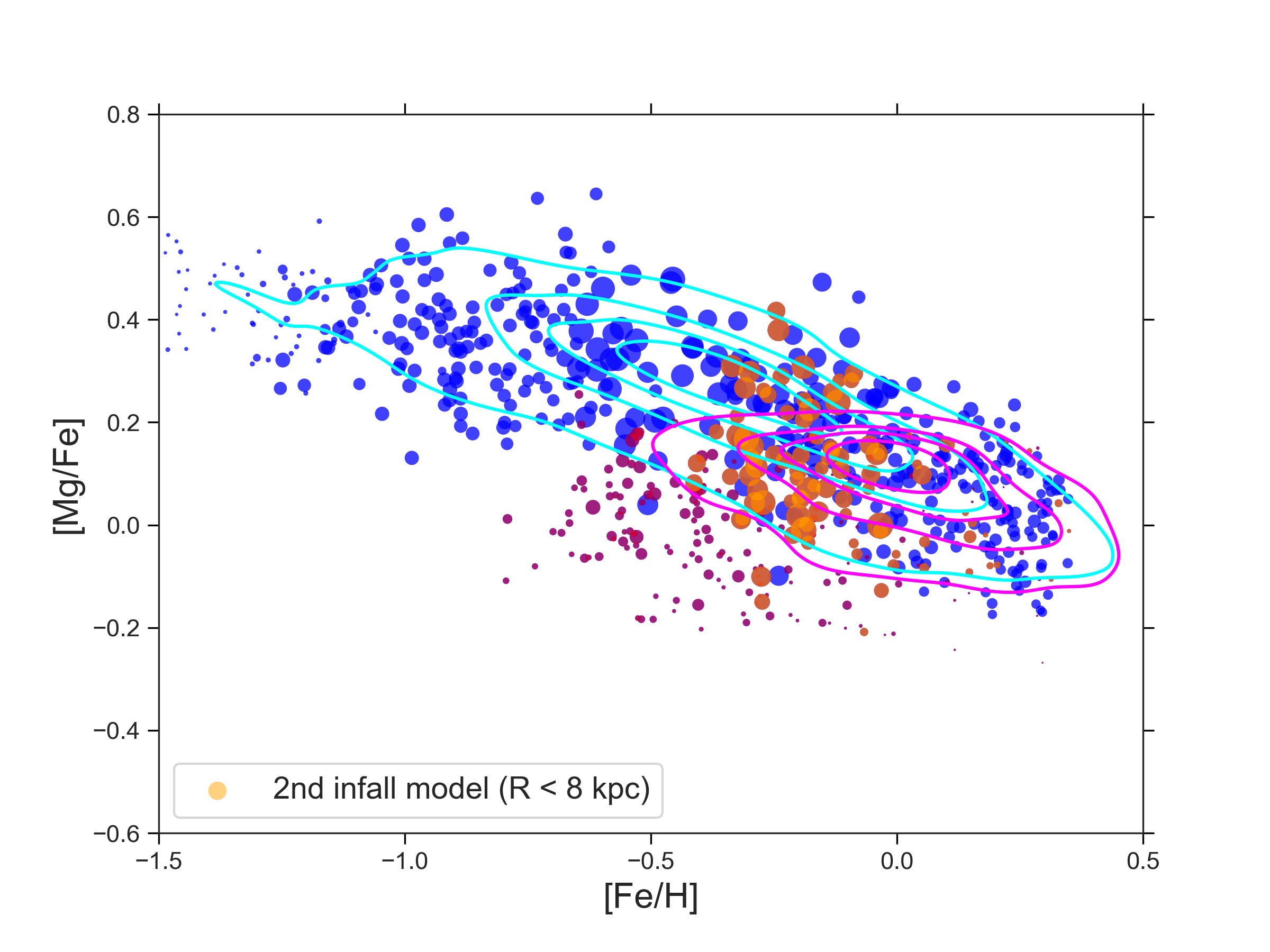}
    \caption{\tiny 
      Two-infall model predictions for [Mg/Fe] vs. [Fe/H]. Left panel: Coloured points are the synthetic model predictions for stars older than 10 Gyr, with orange points associated to the second gas infall episode, while blue points to the first infall. Point size is proportional to the number of stars formed in the model timestep. Data are as in Fig. \ref{fig:model}. Right panel: same as left panel, but with synthetic model predictions associated to stars with R$<$8 kpc coloured in orange.}
    \label{fig:model2}

\end{figure*}

However, additional considerations must be drawn. In the usual picture, the two gas accretion episodes in the two-infall framework are associated to the high-$\alpha$ and low-$\alpha$ sequences, respectively. However, this “classical” picture is altered in the presence of stellar migration for ages $\sim$10 Gyr, especially for the contribution of the inner disk regions. Indeed, the bursty star formation in these regions of the disk, together with the possible infall pre-enrichment (see \citealt{Palla20,Spitoni21} for more details), allow to form high-$\alpha$ stars even after the beginning of the second infall episode. This is shown in Fig. \ref{fig:model2}, where we label model mock data associated to the first and second infall episodes. It can be clearly seen that also predicted stars associated to the second gas accretion episode can be associated to high-$\alpha$ stars, with such stars likely migrated from the inner regions. To better highlight this feature, in Fig. \ref{fig:model2} right panel we highlight with a different color predicted mock stars coming from regions inner than the solar radius (R$\simeq$ 8 kpc). 
In addition, Fig. \ref{fig:model2} also shows that mock data associated with the first infall episode can be identified with low-$\alpha$ data, with the tail of the first star formation episode producing such low-$\alpha$, high-metallicity ([Fe/H]$\gtrsim$ 0 dex) stars.
It is worth noting that such combined features can explain the presence of both low-$\alpha$ and high-$\alpha$ stars across different time bins (see Fig. \ref{fig:MgFe} for Mg), considering i) the smooth transition towards lower [$\alpha$/Fe] and larger [Fe/H] during the first infall and ii) the simultaneous presence of both high-$\alpha$ and low-$\alpha$ stars after the onset of the second infall, as due to the mixing of stars born in different (radial) regions of the disk.

The adoption of such detailed model of Galactic chemical evolution gives already important pieces of information on the history of star formation of the early Galactic disk, proposing a possible scenario for the chemical evolution of the old MW disk and 
highlighting the crucial role covered by stellar radial migration in the Galaxy \citep[e.g.][]{Minchev18,Ratcliffe23,Chen26}. 
Nonetheless, more in-depth analysis is certainly needed to better investigate the chemical enrichment of the analysed stellar sample, including a comparison between different proposed frameworks in the context of Galactic chemical enrichment \citep[e.g.][]{Sharma22,grisoni2026arXiv260510596G}. This will be done in future work, possibly taking advantage of spectroscopic follow-up within the sample, exploiting several classes of chemical elements from different enrichment processes on different timescales.

\section{Summary and conclusions}
\label{sec:conclusion}

Building on the GALAH DR4 catalogue \citep{buder2025PASA...42...51B}, we  characterised a sample of old subgiant-branch (SGB) stars with reliable age estimates, typically precise to 0.5--0.8~Gyr, based on accurate \teff\ determinations and distances from {\it Gaia} parallaxes. We find that stars with solar-like metallicities and low [$\alpha$/Fe] are present up to $\approx$12.5~Gyr, confirming the existence of an old $\alpha$-poor population at metallicities characteristic of the solar neighbourhood. The fractions of high- and low-[$\alpha$/Fe] stars remain comparable across the 8--12.5~Gyr age range, indicating that the two populations overlap substantially in age. The old low-[$\alpha$/Fe] stars also exhibit orbital properties compatible with those of younger $\alpha$-poor stars typically associated with the thin disc.

We provide a catalogue of stars older than 10~Gyr selected for spectroscopic follow-up. These targets are sufficiently bright ($V<14$~mag) for moderate-quality spectroscopy.
Given their ages, metallicities, and location in the SGB region, a subset of these stars are likely to be old solar analogues, i.e. stars that formed with atmospheric properties broadly similar to those of a newborn Sun and have since evolved away from their zero-age main-sequence positions.
Detailed spectroscopic characterisation of these targets will allow us to test their chemical composition and investigate whether solar-like stellar environments were already present during the early stages of thin-disc formation.

Together, these results establish the presence of a previously underexplored old $\alpha$-poor solar-metallicity population and provide a well-defined set of targets for further spectroscopic investigation, opening new avenues for constraining the formation and evolution of the Galactic thin disc. 
In particular, the identification of candidate old solar analogues provides a direct observational pathway to investigate how early solar-like chemical conditions were established in the Milky Way.
The stars selected in this work, together with their ages and other stellar parameters, are publicly available through Zenodo\footnote{\url{https://doi.org/10.5281/zenodo.21208015}}.

We compared the observed population with detailed chemical evolution models for the MW disc \citep{Palla22}, finding that $\alpha$-poor stars can indeed be very old and coeval to $\alpha$-rich population in the tested evolutionary framework. Nonetheless, more detailed chemical analyses and comparison with diverse scenarios for Galactic evolutionary models are required to fully characterise the sample nucleosynthetic history and to understand their role in the early evolution of the Galactic disc.

\begin{acknowledgements}
    The authors thank V. D'Orazi for discussions and support to the manuscript.
    The authors thank Jose Schiappacasse Ulloa for discussions and support with data collection. 
    The authors thank the anonymous referees for constructive feedback that improved the clarity of the paper. R.E.G. and L.M. acknowledge support from INAF through the Large Grants EPOCH and WST, funding for the WEAVE project, the Mini-Grants Checs (1.05.23.04.02), and financial support under the National Recovery and Resilience Plan (PNRR), Mission 4, Component 2, Investment 1.1, Call for tender No. 104 published on 2 February 2022 by the Italian Ministry of University and Research (MUR), funded by the European Union – NextGenerationEU, through the Project ‘Cosmic POT’ (Grant Assignment Decree No. 2022X4TM3H, MUR). M.P. acknowledges support from HORIZON-INFRA-2024-DEV-01-01 – Research Infrastructure Concept Development, through the project WST: The Wide-Field Spectroscopic Telescope (Grant No. 101183153).
    Use was made of the Simbad database, operated at the CDS, Strasbourg, France, and of NASA’s Astrophysics Data System Bibliographic Services. 
    This research used Astropy,\footnote{http://www.astropy.org} a community-developed core Python package for Astronomy \citep{astropy:2018}.
    This work presents results from the European Space Agency (ESA)
    space mission {\it Gaia}. {\it Gaia} data are processed by the {\it Gaia} Data Processing and Analysis Consortium (DPAC). Funding for the DPAC is provided by national institutions, in particular the institutions participating in the {\it Gaia} MultiLateral Agreement (MLA). The {\it Gaia} mission website is \url{https://www.cosmos.esa.int/gaia}. The Gaia archive website is \url{https://archives.esac.esa.int/gaia}.
\end{acknowledgements}

\bibliographystyle{aa.bst}

\bibliography{Faint2}

\begin{appendix} 



    
\section{Compatibility between parameter scales}
\label{sec:compat_params}
\begin{figure}
    \centering
    \includegraphics[width=0.49\linewidth]{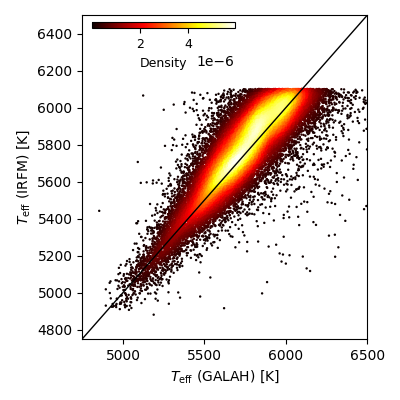}
    \includegraphics[width=0.49\linewidth]{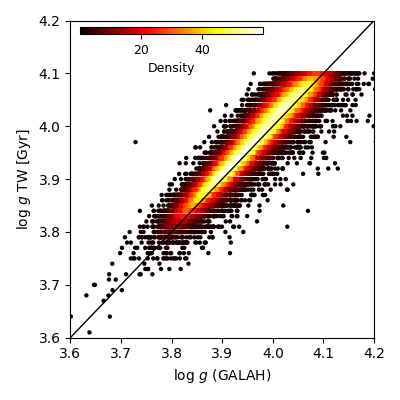}
    
    \caption{\tiny 
     Comparison of \teff\ and \logg\ scales.
     The dispersion in the plots correspond to the {\it characterised sample} selected in Sect.~\ref{sec:sel_old}.
     }
    \label{fig:teff_comp}
\end{figure}

Left panel in Fig.~\ref{fig:teff_comp}
shows the compatibility between the IRFM \teff\ scale of GALAH~DR3 \citep[][used in this work]{casagrande2021}  and the spectroscopic \teff\ scale of GALAH~DR4. The latter is the base of the chemical abundances used in this work. 
Right panel compares the \logg\ scale derived in this work (related to our age estimates according to Sect.~\ref{sec:ages}), with that of GALAH~DR4.

\section{Age validation with the cluster NGC 6352}
\label{sec:validation}

NGC~6352 is an old globular cluster with an age of 10--12.5~Gyr, as inferred from isochrone analyses by \citet{vandenberg2013ApJ...775..134V} and \citet{oliveira2020ApJ...891...37O}. It has a metallicity of $\mathrm{[Fe/H]} = -0.55 \pm 0.03$~dex \citep{feltzing2009A&A...493..913F} and follows a thick-disc-like orbit \citep{perez2020MNRAS.491.3251P}. To validate our methodology, we performed both a global isochrone fit in the $M_V$--\teff\ plane and individual star-by-star age determinations.

Our analysis is based on the observational compilation of \citet{vasilev2021MNRAS.505.5978V}. Starting from a sample of 3408 stars with Gaia photometry and membership probabilities exceeding 90\%, we retained only sources with uncertainties in both Gaia $G_{BP}$ and $G_{RP}$ magnitudes below 0.05~mag. This quality criterion reduced the sample to 2113 stars, of which 751 have membership probabilities above 99\%.
Photometric \teff\ were derived from the  $G_{BP}-G_{RP}$ colour using the calibrations of \citet{casagrande2021}, which are tied to the IRFM temperature scale established from GALAH~DR3 stars. We adopted $G_{BP}-G_{RP}$ in preference to colours involving the $G$ band in order to minimise systematic effects in the $G-G_{BP}$ and $G-G_{RP}$ colour scales \citep{giribaldi2023A&A...679A.110G}. 
The calibrations require \logg, [Fe/H], and reddening $E(B-V)$ as inputs. We adopted fixed values of $\mathrm{[Fe/H]} = -0.55 \pm 0.10$~dex and $E(B-V) = 0.24 \pm 0.048$~mag from \citet{feltzing2009A&A...493..913F}. The associated uncertainties are needed for the
analysis of the errors. The  [Fe/H] error reflects  typical values in the literature, while the reddening uncertainty was assumed to be of 20\%.
$E(B-V)$ was transformed to 
into the reddening $E(G_{BP} - G_{RP})$ and the attenuation $A_G$ to derive $V_0$ via Eq.~\ref{eq:gaia_johnson} assuming the COD extinction law, as explained in Appendix~\ref{sec:redd}.
Absolute magnitude $M_V$ was computed as $M_V = V_0 - 5 \rm{log} (d) + 5$, where $\rm{d} = 5543 \pm 73$~pc is the cluster distance \citep{Baumgardt2021MNRAS.505.5957B}. 

\begin{figure}
    \centering    \includegraphics[width=1\linewidth]{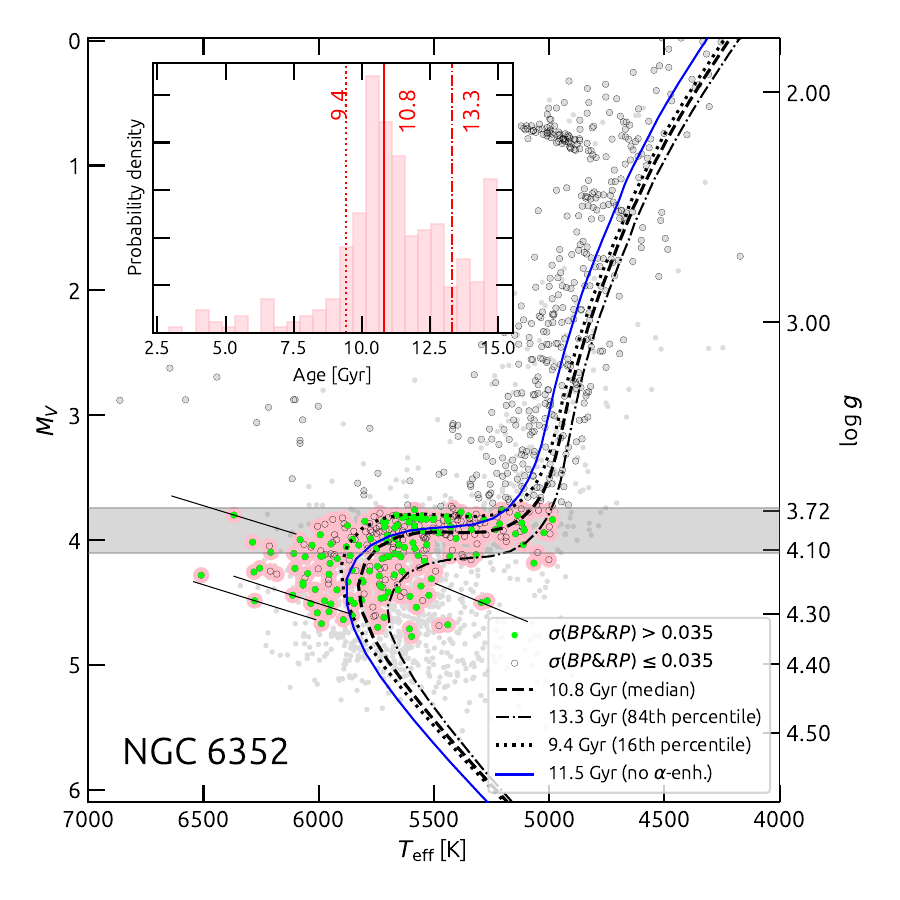}
    \caption{\tiny 
     $M_V$--\teff\ diagram of NGC~6352. Grey and black  circles represent cluster stars with membership probabilities higher than 90\% and 99\%, respectively. Turnoff stars used for the isochrone fitting are highlighted in pink.
     Stars with photometric precision $\sigma(G_{BP})$ and $\sigma(G_{RP})$ between 0.04 and 0.05 are shown in green.
     Black lines represent isochrones with ages corresponding to the 16th, median, and 84th quantiles of the histogram. 
     An isochrone with no $\alpha$-enhancement effect on [Fe/H] is displayed in blue.
     Oblique lines exemplify changes of positions of a few stars with the most imprecise photometry.
     The right axis displays the \logg\ scale of  isochrone corresponding to the median age.
     The grey band indicates the subgiant branch at $3.76 < M_V < 4.10$~mag.
     The inset displays the histogram of the star-by-star age determination. Its median (solid line) and 16th and 84th percentiles (dotted lines and dash-dotted lines) are noted in red.     
     }    
     \label{fig:iso_fit}
\end{figure}

Initial \logg\ estimates were assigned according to Gaia $G$ magnitude: \logg\ = 4 for $G > 18$, \logg\ = 3 for $16 < G \leq 18$, \logg\ = 2 for $14 < G \leq 16$, \logg\ = 1 for $12 < G \leq 14$, and \logg\ = 0.5 for $G \leq 12$. 
Uncertainties in these preliminary values have only a minor impact on the derived temperatures; for example, an error of 0.5~dex in \logg\ changes \teff\ by less than 50~K. Refined temperatures were therefore obtained through a second iteration using updated surface gravities.
Surface gravities and individual stellar ages (star-to-star henceforth) were derived following the procedure described in Sect.~\ref{sec:ages}. Two iterations were performed, with the second adopting the revised \teff\ values. An $\alpha$-enhancement of [$\alpha$/Fe]$=+0.2$~dex \citep{feltzing2009A&A...493..913F} was assumed via Eq~\ref{eq:alpha_en}.

This validation focuses exclusively on turnoff stars; therefore, we restricted the sample to $17.45 < V_0 < 18.60$ and performed isochrone fitting using a method similar to that described by \citet{giribaldi2023A&A...673A..18G}. This selection yielded  293 stars  with $3.76 < M_V < 4.79$~mag.
Figure~\ref{fig:iso_fit} presents the $M_V$--\teff\ diagram for NGC~6352. Stars with membership probabilities exceeding 99\% are shown in black, while the turn-off sample is highlighted in pink. 
Stars with the most imprecise photometry are shown in green.
The dashed line represents the isochrone corresponding to the median of the star-to-star ages, 10.8~Gyr, which is in excellent agreement with the value of $10.75 \pm 0.78$~Gyr reported by \citet{vandenberg2013ApJ...775..134V}.
For comparison, an isochrone computed without $\alpha$-enhancement effect on [Fe/H] is also shown (blue curve). Its close proximity to the best-fitting solution indicates that neglecting [$\alpha$/Fe] would lead to an age overestimate of 1~Gyr.

\section{Sources of age uncertainties and biases estimated from  analysis of the NGC 6352 cluster}
\label{sec:deviation}
Oblique vectors in Fig.~\ref{fig:iso_fit} illustrate the displacements from the combined uncertainties in reddening, photometry, and the colour--\teff\ calibration. 
These correspond to \teff\ uncertainties of 200-250~K for stars with typical photometric precision ($\sigma(G_{BP})=0.029$~mag and $\sigma(G_{RP})=0.022$~mag) and up to 280~K for those near the adopted quality threshold ($\sigma(G_{BP})=\sigma(G_{RP})=0.05$~mag). 
The \teff\ uncertainties include contributions from the intrinsic scatter of the colour--\teff\ calibration \citep[60~K, Table~1 of ][]{casagrande2021} and reddening uncertainty ($\sim$200~K).
The uncertainty in the absolute magnitude, $\sigma(M_V) \simeq 0.15$~mag, is also dominated by the reddening error. It was estimated by adding in quadrature the following sources: (i) the scatter of the Gaia--Johnson colour transformations, $\sigma(V_{calib})=0.03$~mag; (ii)  the uncertainty of the dereddened magnitude $\sigma(V_0) = 0.148\, \mathrm{mag} = 3.1 \times 0.2 \times E(B-V)$, where 0.2 is the 20\% error of reddening assumed and 3.1 is the extinction coefficient, and (iii)  the cluster distance uncertainty (73~pc), which translates into $0.029$~mag through the distance modulus.

The age distribution in Fig.~\ref{fig:iso_fit} exhibits an extended tail toward older ages, with the 16th and 84th percentiles corresponding to dispersions of 1.4 and 2.5~Gyr about the median, respectively.
This asymmetry contrasts with the distribution of temperature residuals (observational \teff\ minus isochrone values) computed for the 10.8~Gyr solution, which is symmetric. Namely, it has 16th, median, and 84th percentiles of $-291$, $-2$, and $+321$~K, and a standard deviation of 304~K. This indicates that stars scattered toward cooler temperatures are preferentially mapped into older ages, producing the observed high-age tail the histogram exhibits.

\begin{figure}
    \centering
    \includegraphics[width=1\linewidth]{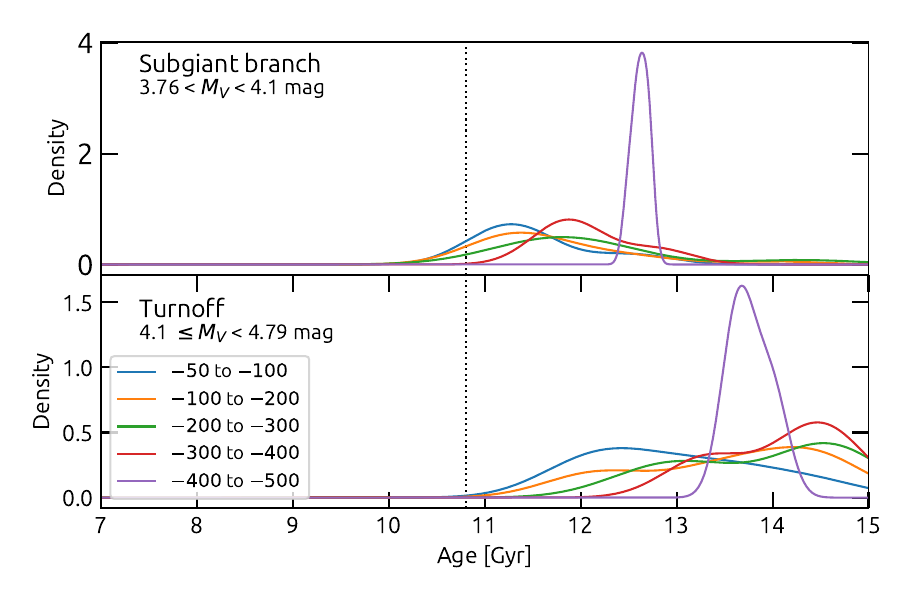}
    
    \caption{\tiny 
     Probability density distributions of stellar ages for cluster stars with negative temperature residuals ($\Delta T_{\rm eff}=T_{\rm eff,obs}-T_{\rm eff,iso}<0$). Stars are grouped according to their temperature residuals, as indicated in the legend, from the least discrepant ($-50$ to $-100$~K) to the most discrepant ($-400$ to $-500$~K). The top and bottom panels show the SGB and TO stars, respectively, separated according to their $M_V$ values as indicated in the labels.
     The vertical dotted line indicates the most frequent age of 10.8~Gyr of the star-to-star estimates in Fig.~\ref{fig:iso_fit}.
     }
    \label{fig:t_dev}
\end{figure}

Figure~\ref{fig:t_dev} identifies the stars contributing to the high-age tail. It shows age density distributions for stars located to the cool side of the 10.8~Gyr isochrone in Fig.~\ref{fig:iso_fit}. The distributions are grouped according to their \teff\ residuals relative to the isochrone, from the least to the most discrepant intervals, following a colour-code.
Additionally, the subgiant branch (SGB) and the turnoff (TO) are separated in the top and bottom panels; a limit of $M_V = 4.1$~mag (gray band in Fig.~\ref{fig:iso_fit}) was imposed.
The SGB exhibits modest age offsets up to $\sim$1.5~Gyr for \teff\ underestimations as large as $-400$~K. 
In contrast, the TO shows substantially larger age biases, even for \teff\ underestimations lower than 100~K. It demonstrates that \teff\ underestimations in the TO exceeding 100~K, systematically populate the high-age tail  of the histogram of Fig.~\ref{fig:iso_fit}, being largely responsible for inferred ages above the age of the universe  13.7~Gyr.

\begin{figure}
    \centering
    \includegraphics[width=1\linewidth]{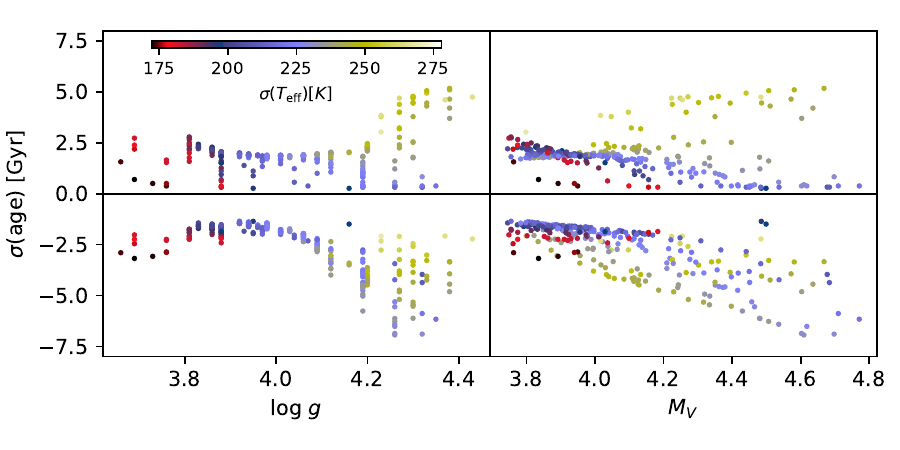}
    
    \caption{\tiny
     Asymmetric age uncertainty of NGC~6352 SGB and TO stars. Values are plotted as functions of \logg\ and $M_V$ in left and right panels. The colour coding indicates the \teff\ uncertainty.
     }
    \label{fig:sigma_age}
\end{figure}

This analysis indicates that the SGB yields  unbiased star-by-star age determinations\footnote{This refers to the application of  the frequentist approach.}, because at this stage the impact of the \teff\ deviations from true values nullifies.
For the cluster NGC~6352, the SGB remains within $3.72 \lesssim \logg \lesssim 4.10$~dex.
This effect is also observed in the errors of the star-by-star ages in Fig.~\ref{fig:sigma_age}, which are minimal (lower than $\sim$3~Gyr) for \logg~$\lesssim 4.1$~dex or $M_V \lesssim 4.1$~mag. 

\begin{figure}
    \centering
    \includegraphics[width=1\linewidth]{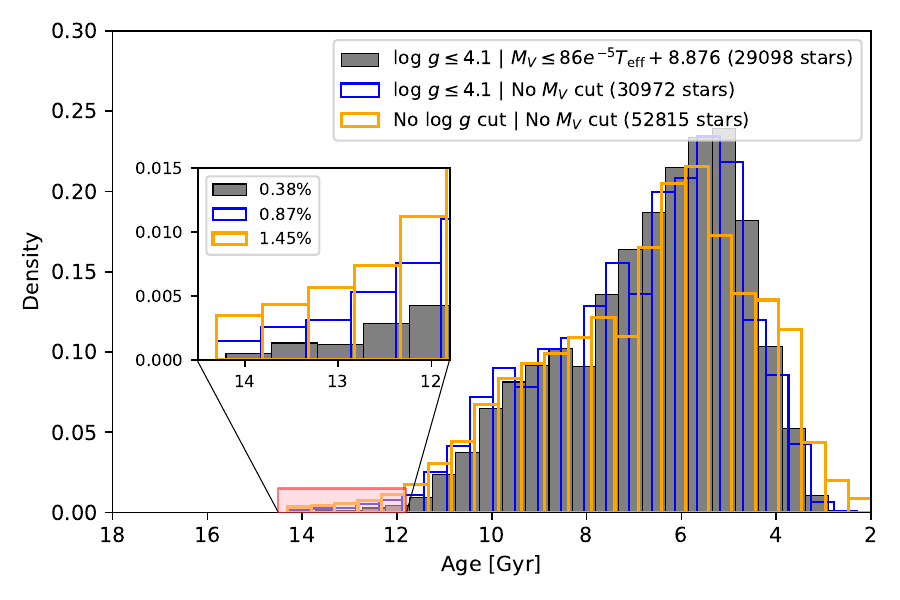}
    
    \caption{\tiny  
     Age normalised histograms.
     The gray, blue, and orange histograms consider the \logg\ and $M_V$ cuts indicated in the legends. The gray histogram is the same as in Fig.~\ref{fig:histo_eye}. 
     The inner plot expands the area shaded in pink.
     The quantities indicate the percentages of stars older than 12~Gyr in each sample. 
     }
    \label{fig:hist_comp}
\end{figure}

The $M_V$ range of the SGB is primarily determined by stellar age and metallicity.
In Sect~\ref{sec:sel_old}, only SGB field stars are present in the sample.
Figure~\ref{fig:hist_comp} compares the histogram shown in Fig.~\ref{fig:histo_eye}  with with those obtained without the \logg\ and $M_V$ cuts.
The application of these cuts results in a progressively smaller fraction of stars with ages above 10~Gyr. In particular, the fraction of stars older than 12~Gyr decreases from 1.45\% in the full sample to 0.87\% after applying the \logg~$<=4.1$~dex cut, and further to 0.38\% after the $M_V$ cut.
This reduction demonstrates that the \logg\ and $M_V$ cuts effectively suppress the contribution of stars whose ages are spuriously overestimated due to small underestimations of \teff.

\section{Assessment of the impact of reddening on the age estimate}
\label{sec:redd}

\begin{figure}
    \centering
    \includegraphics[width=1\linewidth]{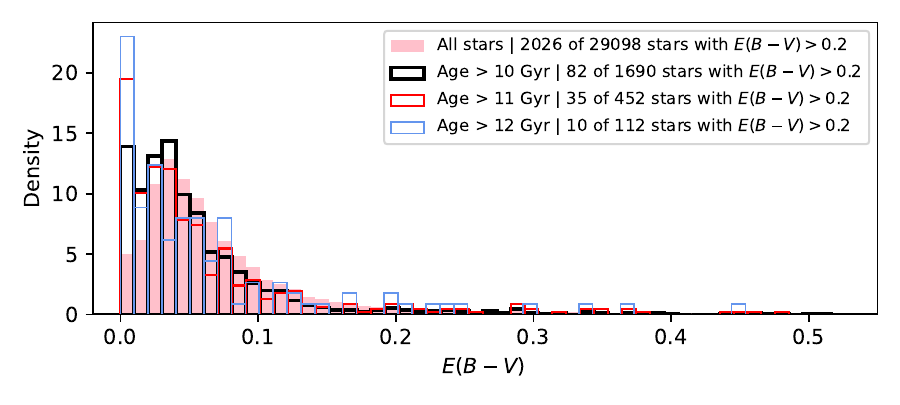}
    
    \caption{\tiny 
      Reddening distributions for the characterised sample and its age-selected subsamples. Different colours denote the full sample and the subsamples defined by the adopted age thresholds, as indicated in the legend. The fraction of stars with $E(B-V) > 0.2$ is given for each sample.
     }
    \label{fig:ebv_charsample}
\end{figure}

The IRFM \teff\ adopted in this work is based on intrinsic band magnitudes, whose extinction corrections follow the extinction law of \citet{cardelli1989ApJ...345..245C} and \citet{odonnell1994ApJ...422..158O}, hereafter COD.
\cite{casagrande2021} showed that the effect of instead adopting the extinction law of \cite{fitzpatrick1999PASP..111...63F} renormalised as per \cite{Schlafly2011ApJ...737..103S} (hereafter FSF), becomes significant (i.e. it yields \teff\ cooler by at least 100~K) for $E(B-V) \gtrsim 0.2$~mag; see Fig.~B1 in the paper. 
Our {\it characterised sample} (Sect.~\ref{sec:sel_old}) has a marginal fraction of stars with $E(B-V)>0.2$~mag, as shown in Fig.~\ref{fig:ebv_charsample}.
Therefore, if the true reddening of our stars lie somewhere in between COD and FSF estimates, the \teff\ of our sample would be biased to hotter, leading to somewhat younger ages (see bias map in Fig.~\ref{fig:map_error}).
Nevertheless, even if it were the case,  these age biases would be limited to a minority of 5-10\%; see fractions in Fig.~\ref{fig:ebv_charsample}.
In any case, since the characterised sample only includes SG stars with \logg~$< 4.1$~dex, our age estimates are almost insensitive to potential \teff\ biases, as the values in Figs.~\ref{fig:map_error} and \ref{fig:iso_fit} show.

Our age estimate therefore is primarily  influenced by  the reddening accuracy via the intrinsic magnitude $V_0$, which was computed from the Gaia $G$, $G_{BP}$, and $G_{RP}$ magnitudes by Eq.~\ref{eq:gaia_johnson} from \citet{riello2021A&A...649A...3R}:
\begin{equation}
\label{eq:gaia_johnson}
\begin{split}
    V = {} & G + 0.02704 - 0.01424 \times (G_{BP} - G_{RP}) \\
           & + 0.2156 \times (G_{BP} - G_{RP})^2 - 0.01426 \times (G_{BP} - G_{RP})^3
\end{split}
\end{equation}

\noindent where each Gaia magnitude band had previously been corrected for extinction.
Such corrections were computed by multiplying $E(B-V)$ by the extinction coefficients $R_G$, $R_{G_{BP}}$, and $R_{G_{RP}}$, derived using the equations of Table~B1 in \cite{casagrande2021} based on the COD extinction law, for consistence with IRFM \teff.
We adopted the reddening values in GALAH~DR3. For the vast majority\footnote{In the characterised sample, only eight out of $29\,098$ stars have reddening ``rjc", as tagged in GALAH~DR3, which refers to the Rayleigh-Jeans colour excess method \citep{majewski2011ApJ...739...25M}.} of our stars, these were compiled by \cite{casagrande2021}, who selected the values of \cite{green2019ApJ...887...93G} (Bayestar hereafter) when available, or from \cite{SFD} (SFD hereafter) re-scaled as in \cite{casagrande2019MNRAS.482.2770C}.

\begin{figure}
    \centering
    \includegraphics[width=1\linewidth]{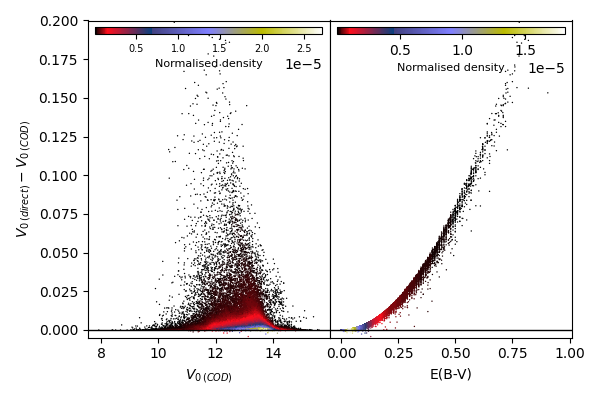}
    
    \caption{\tiny 
      Bias of $V_0$ from correcting extinction on Gaia-transformed $V$.
     Values are relative to $V_0$ derived from $G_0$ and $(G_{BP} - G_{RP})_0$ in Eq.~\ref{eq:gaia_johnson} assuming the COD extinction law.
     Left and right panels show values as function of $V_0$ and reddening, respectively.
     The values exhibited correspond to the sample of $91\,425$ stars with \teff\ $< 6100$~K (See Sect.~\ref{sec:sel_old}).
     }
    \label{fig:reddd2}
\end{figure}

The equations for deriving the extinction coefficients of \cite{casagrande2021} are calibrated as function of the intrinsic (i.e. reddening-corrected) colour $(G_{BP} - G_{RP})_0$.
As recommended by the authors, for optimal results, the extinction coefficients were computed iterating the calculations, starting from $(G_{BP} - G_{RP})_0 \simeq (G_{BP} - G_{RP}) - E(B-V)$; we iterated ten times. 
We note that computing $V_0$ from an intuitive direct application of the relation $V - 3.1 \times E(B-V)$ instead of using Eq.~\ref{eq:gaia_johnson} with extinction-corrected magnitudes carries biases that grow with $E(B-V)$, as exhibited in Fig.~\ref{fig:reddd2}.

\begin{figure}
    \centering
    \includegraphics[width=1\linewidth]{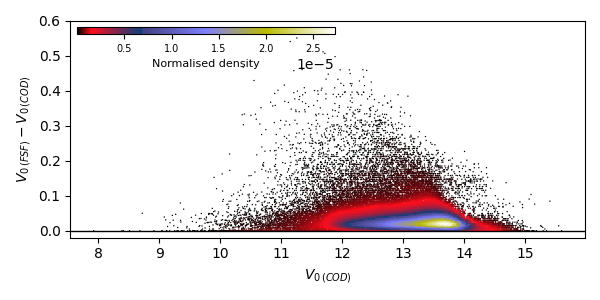}
    \includegraphics[width=1\linewidth]{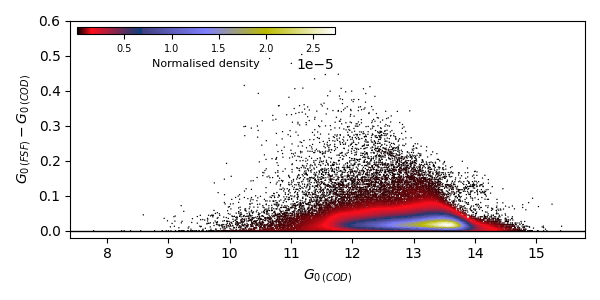}
    \includegraphics[width=1\linewidth]{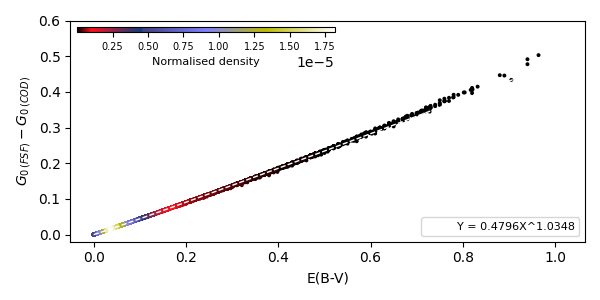}
    \caption{\tiny  
     Difference between intrinsic magnitudes computed from FSF and COD reddening laws. {\it Top and middle panels: } The concentration of the distributions are colour-coded according to the bars. 
     {\it Bottom panel:} Difference of intrinsic $G_0$ magnitudes as function of reddening. 
     The concentration of the stars is mapped along the horizontal axis and colour-codded according to the bar. The trend is fitted by the white dashed line with a power-law with parameters noted in the plot.
     The values exhibited correspond to the same sample used in Fig~\ref{fig:reddd2}
     }
    \label{fig:reddd}
\end{figure}

The top and middle panels of Fig.~\ref{fig:reddd} show the differences in $V_0$ and $G_0$, respectively, between the values computed using the FSF and COD extinction laws. Their nearly identical dispersions indicate that the differences in $V_0$ are driven almost entirely by those in $G_0$ through Eq.~\ref{eq:gaia_johnson}. The bottom panel shows that these differences correlate with $E(B-V)$ and follow a power-law relation, with the FSF prescription yielding systematically higher magnitudes. For most stars, the difference in $G_0$ is smaller than 0.1~mag.
According to the quantities in Fig.~\ref{fig:map_error}, adopting the FSF prescription instead of COD would result in a positive shift in the inferred ages of SGB stars, with a maximum amplitude of 1~Gyr for $E(B-V)<0.2$~mag.

\begin{figure}
    \centering
    \includegraphics[width=1\linewidth]{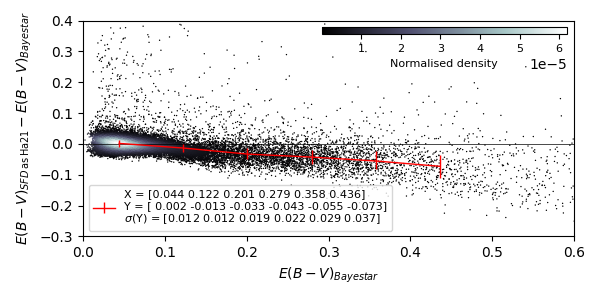}
    
    \caption{\tiny 
     Difference between SFD-based and Bayestar $E(B-V)$ values as function of $E(B-V)$. 
     Red lines connect medians computed in equally spaced bins, and vertical bars correspond to 1$\sigma$ dispersions, their values are indicated in the plot.
     The values represented by individual points correspond to the sample in Figs.~\ref{fig:reddd2} and \ref{fig:reddd} with data available in the Bayestar and SFD maps, and  with galactic latitude $|b| > 10^{\circ}$; this is composed of $35\, 758$ stars.
     }
    \label{fig:red_disp}
\end{figure}

For estimating the precision of the GALAH~DR3 reddening, required for the estimate of the age error budget, we compared the SFD and Bayestar reddening values extracting the data through the \texttt{dustmaps} package \citep{green2018JOSS....3..695G}.
SFD was re-scaled by Eq.~1 of \cite{hansen2021MNRAS.501.5309H}.
Figure~\ref{fig:red_disp} shows that the offset between the two scales and the dispersion increase with $E(B-V)$, as traced by the red line sequence, according to the quantities in the plot legends. The consistency holds within 1$\sigma$ up to $E(B-V) \sim 0.2$~mag.
We attributed half of the dispersion to each scale, the corresponding  values of which were thus fitted linearly as function of $E(B-V)$ to obtain the reddening uncertainty as expressed below

\begin{equation}
\label{eq:reddening_error}
\sigma(E(B-V)) = 0.033 \times E(B-V) + 0.003.
\end{equation}

The origin of the systematic offset between the two reddening scales is beyond the scope of this work. We note, however, that, apart from a few outliers, the oldest stars in the characterised sample have $E(B-V)$ values within the range where the two scales agree (Fig.~\ref{fig:ebv_charsample}), indicating that their derived ages are largely insensitive to potential reddening systematics.

\section{Comparison with catalogued ages}
\label{sec:ages_C}

\begin{figure}[!htbp]
    \centering
    \includegraphics[width=0.49\linewidth]{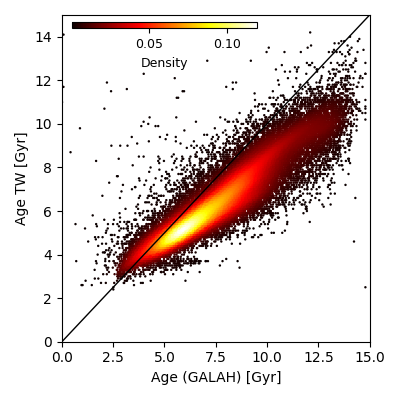}
    \includegraphics[width=0.49\linewidth]{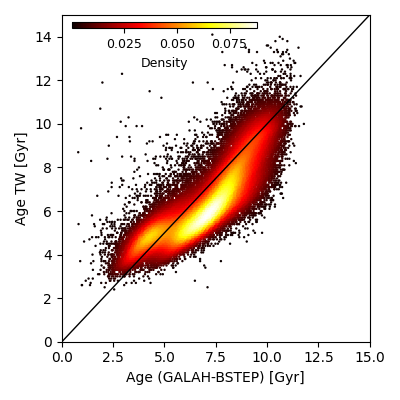}
    
    \caption{\tiny 
     Comparison of stellar ages. The plots contain the characterised sample, the age histogram of which shown in Fig.~\ref{fig:histo_eye}.
     Left and right panels show the `on the fly' and `BSTEP' ages of the GALAH catalogue in the horizontal axis, respectively. 
     }
    \label{fig:ages_C}
\end{figure}

Figure~\ref{fig:ages_C} compares our age estimates with the
GALAH {\it on-the-fly} ages, derived through frequentist inference, and the GALAH BSTEP ages, derived through Bayesian inference \citep{sharma2018MNRAS.473.2004S}. Both GALAH age estimates are based on PARSEC+COLIBRI isochrones \citep{bressan2012MNRAS.427..127B,marigo2017ApJ...835...77M}. For the range age $\gtrsim 5$~Gyr, both GALAH age scales exceed our estimates by approximately 2--3~Gyr. The BSTEP ages appear truncated above at 12~Gyr. It is possible that the origin of the discrepancies remain in the different set of isochrones used. However, discrepancies may arise from the diverse details involved in the implementations of the methods and the specific adopted atmospheric, photometric, and astrometric parameters. Therefore, an answer requires a differential analysis.
In what regards the age range of interest of this work (> 10~Gyr), our ages and the BSTEP ones seem to find better agreement. 
However, many stars considered `young' in our sample (i.e. of 5 to 10~Gyr) are found to be older than 10~Gyr by the BSTEP procedure.

\end{appendix}

\end{document}